\documentclass[twocolumn]{aastex62}

\usepackage{graphicx}
\usepackage{float}
\usepackage{mathtools}
\usepackage{color}
\usepackage{tabularx}
\usepackage{sidecap}
\usepackage[caption = false]{subfig}

\usepackage{longtable}
\usepackage{acronym}
\usepackage[flushleft]{threeparttable}
\usepackage{hyperref}

\def\lesssim{\mathrel{\hbox{\rlap{\hbox{\lower4pt\hbox{$\sim$}}}\hbox{$<$}}}}
\def\gtrsim{\mathrel{\hbox{\rlap{\hbox{\lower4pt\hbox{$\sim$}}}\hbox{$>$}}}}

\acrodef{DLT40}[DLT40]{Distance Less Than 40 Mpc survey}
\acrodef{CBC}[CBC]{compact binary coalescence}
\acrodef{GW}[GW]{gravitational wave}
\acrodef{NS}[NS]{neutron star}
\acrodef{EM}[EM]{electromagnetic}
\acrodef{GRB}[GRB]{Gamma Ray Burst}
\acrodef{SN}[SNe]{Supernovae Type Ia}
\acrodef{LVC}[LVC]{Laser Interferometer Gravitational-wave Observatory and Virgo collaboration}
\acrodef{BNS}[BNS]{binary neutron star}
\acrodef{BBH}[BBH]{binary black hole}
\acrodef{NSBH}[NSBH]{black hole neutron star merger}

\received{XXX}
\revised{YYY}
\accepted{ZZZ}
\submitjournal{ApJ}

\shorttitle{Electromagnetic Followup of LIGO events with DLT40}
\shortauthors{Yang et al.}

\begin{document}

\title{Optical follow-up of gravitational wave events during the second advanced LIGO/VIRGO observing run \\ with the DLT40 Survey}

\correspondingauthor{Sheng Yang}
\email{sheng.yang@inaf.it}

\author[0000-0002-2898-6532]{Sheng Yang}
\affil{Department of Physics, University of California, 1 Shields Avenue, Davis, CA 95616-5270, USA}
\affil{Department of Physics and Astronomy Galileo Galilei, University of Padova, Vicolo dell'Osservatorio, 3, I-35122 Padova, Italy}
\affil{INAF Osservatorio Astronomico di Padova, Vicolo dell’Osservatorio 5, I-35122 Padova, Italy}

\author[0000-0003-4102-380X]{David J. Sand}
\affiliation{Department of Astronomy/Steward Observatory, 933 North Cherry Avenue, Room N204, Tucson, AZ 85721-0065, USA}

\author[0000-0001-8818-0795]{Stefano Valenti}
\affiliation{Department of Physics, University of California, 1 Shields Avenue, Davis, CA 95616-5270, USA}

\author[0000-0001-5008-8619]{Enrico Cappellaro}
\affiliation{INAF Osservatorio Astronomico di Padova, Vicolo dell’Osservatorio 5, I-35122 Padova, Italy}

\author[0000-0003-3433-1492]{Leonardo Tartaglia}
\affiliation{Department of Astronomy and The Oskar Klein Centre, AlbaNova University Center, Stockholm University, SE-106 91 Stockholm, Sweden}
%\affiliation{Department of Physics, University of California, 1 Shields Avenue, Davis, CA 95616-5270, USA}
\affiliation{Department of Astronomy/Steward Observatory, 933 North Cherry Avenue, Room N204, Tucson, AZ 85721-0065, USA}

\author{Samuel Wyatt}
\affiliation{Department of Astronomy/Steward Observatory, 933 North Cherry Avenue, Room N204, Tucson, AZ 85721-0065, USA}

\author[0000-0003-3433-1492]{Alessandra Corsi}
\affiliation{Physics $\&$ Astronomy Department, Texas Tech University, Lubbock, TX 79409, USA}

\author[0000-0002-5060-3673]{Daniel E. Reichart}
\affiliation{Department of Physics and Astronomy, University of North Carolina at Chapel Hill, Chapel Hill, NC 27599, USA}

\author{Joshua Haislip}
\affiliation{Department of Physics and Astronomy, University of North Carolina at Chapel Hill, Chapel Hill, NC 27599, USA}

\author{Vladimir Kouprianov}
\affiliation{Department of Physics and Astronomy, University of North Carolina at Chapel Hill, Chapel Hill, NC 27599, USA}
\affiliation{Central (Pulkovo) Observatory of Russian Academy of Sciences, 196140 Pulkovskoye Ave. 65/1, Saint Petersburg, Russia}

\collaboration{(DLT40 collaboration)}

\begin{abstract}
We describe the \ac{GW} follow-up strategy and subsequent results of the \ac{DLT40} during the second science run (O2) of the \ac{LVC}.  Depending on the information provided in the \ac{GW} alert together with the localization map sent by the \ac{LVC}, \ac{DLT40} would respond promptly to image the corresponding galaxies selected by our ranking algorithm in order to search for possible \ac{EM} counterparts in real time. During the \ac{LVC} O2 run, \ac{DLT40} followed ten \ac{GW} triggers, observing between $\sim$20-100 galaxies within the \ac{GW} localization area of each event.  From this campaign,  we identified two real transient sources within the GW localizations with an appropriate on-source time -- one was an unrelated type Ia supernova (SN~2017cbv), and the other was the optical kilonova, AT 2017fgo/SSS17a/DLT17ck, associated with the binary neutron star coalescence GW170817 (a.k.a gamma-ray burst GRB170817A).  We conclude with a discussion of the DLT40 survey's plans for the upcoming LVC O3 run, which include expanding our galaxy search fields out to $D$$\approx$65 Mpc to match the LVC's planned three-detector sensitivity for binary neutron star mergers.
\end{abstract}


\section{Introduction}
\label{parintroduction}

The long-awaited direct detection of gravitational waves (GWs) by first the Advanced Laser Interferometer Gravitational-wave Observatory \citep[LIGO; ][]{ligo} and then the Advanced Virgo interferometer \citep{virgo} has ushered in a new era of physics. These exquisite detectors are designed to probe high frequency ($\sim$10--1000 Hz) GW signals whose main astrophysical sources are compact binary coalescences (CBCs) -- i.e. \ac{BBH}, black hole -- neutron star and binary neutron star (BNS) mergers.

During the first Advanced LIGO observing run (O1), two BBH GW events were detected (GW150914; \citealt{GW150914}, and GW151226; \citealt{GW151226}), along with a third likely event of astrophysical origin (LVT151012; \citealt{LVT151012}).  While BBH mergers are not generally expected to have electromagnetic (EM) counterparts\footnote{Whether  a BBH merger can produce an \ac{EM} signal is still an open question. A weak gamma-ray transient was reported 0.4 s after GW150914, consistent with the GW localization \citep{fermi_bbhem}.  Subsequent models and theoretical work \citep{Loeb16,Perna16,Perna18,Zhang16,Bartos17,dk17} make the continued search for EM counterparts to BBHs a source of interest.}, the astronomical community (joined by neutrino and cosmic-ray researchers) promptly responded and searched for transients associated with these BBH events \citep[see e.g.][for a summary of the follow-up of GW150914]{GW150914_EM}, partly in preparation for GW events which are expected to have an EM counterpart, namely those with a neutron star as one of the constituents.

The participation of the EM, neutrino and cosmic ray communities continued in the second LVC observing season (O2), with hundreds of telescopes and detectors  participating in the follow-up of GW sources. During O2, the Advanced LIGO detectors were significantly more sensitive, and were joined by Virgo in the final month of data taking -- this is significant, as the addition of a third detector can decrease the GW localization regions from $\sim$100-1000 deg$^2$ to $\sim$10-100 deg$^2$ for high signal-to-noise events \citep{lvknew}.  In total, fourteen GW alerts were issued by the LVC, six of which were ultimately determined to be true GW events \citep{lvc_19}.  Many of these were aggressively pursued by the astronomical community, whose efforts finally bore fruit with the GW170817 BNS merger \citep{GW170817}, which led to the discovery of the first EM counterpart of a GW event \citep[from gamma-ray to radio wavelengths;][]{GW170817MMA,sss17a,dlt17ck,vista,master,decam,lcogtkn,Fermi_KN,INTEGRAL_KN,Haggard_17,Troja_17,Hallinan_17,Margutti_17}.

In this paper we describe the observational campaign performed by the Distance Less Than 40 Mpc (DLT40) project and its \ac{GW} follow-up efforts during the LVC O2 run.  
Elsewhere we have described the co-discovery of AT2017gfo/SSS17a/DLT17ck and the DLT40 team's follow-up observations \citep{dlt17ck} and the limit to the Kilonova rate using the DLT40 search \citep{dlt40rate}.  
In Section~\ref{sec:DLT40} we provide an overview of the \ac{DLT40} supernova and transient program. A description of the \ac{DLT40} \ac{GW} follow-up strategy in particular is described in Section~\ref{sec:GWfollow}. Next, in Section \ref{sec:O2} we present the results of the \ac{DLT40} follow-up during O2.  We discuss the DLT40 program's specific upgrades and plans for the imminent O3 observing run in Section~\ref{sec:future}.  We end the paper with a brief discussion and summary (Section \ref{sec:discussion}). 

\section{DLT40 transient search}
\label{sec:DLT40}
\begin{figure*}
    \subfloat[The distance distribution of \ac{DLT40} (filled histogram) and GWGC galaxies within 40 Mpc (shaded histogram). 
    \label{subfig-1:dlt40cat}]{%
     \includegraphics[width=0.45\textwidth]{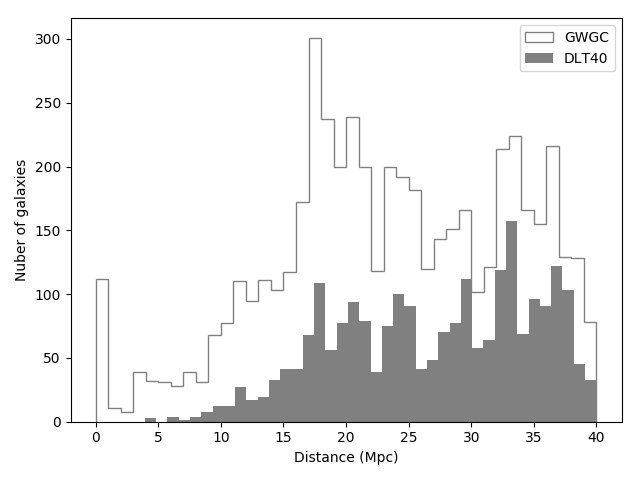}}
     \hfill
     \subfloat[The Hubble type distribution for \ac{DLT40} (shaded histogram) and GWGC galaxies (open histogram) within 40 Mpc. \label{subfig-2:dlt40cat}]{%
       \includegraphics[width=0.45\textwidth]{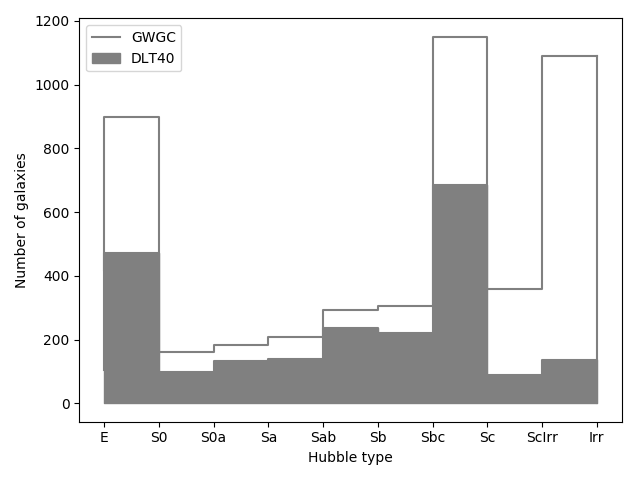}}
     \hfill
     \subfloat[The major axis distribution (in arcmin) for \ac{DLT40} (shaded histogram) and GWGC galaxies within 40 Mpc (open histogram).
     \label{subfig-3:dlt40cat}]{%
       \includegraphics[width=0.45\textwidth]{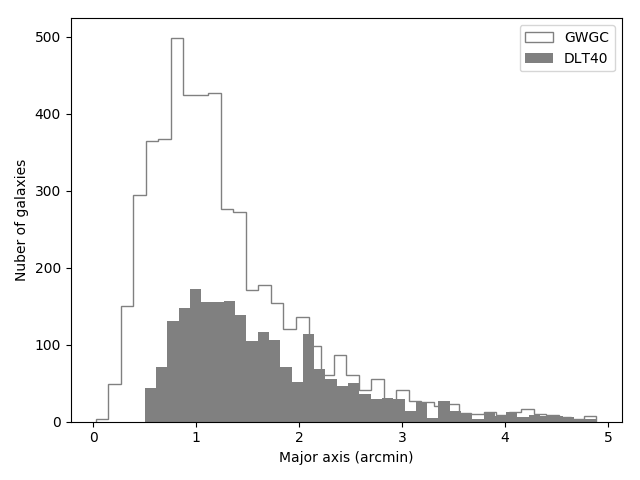}}
     \hfill
     \subfloat[The binned and integrated B band luminosity distribution for \ac{DLT40} and GWGC galaxies within 40 Mpc; note the binned points are difficult to distinguish on the scale of this plot.  
     \label{subfig-5:dlt40cat}]{%
       \includegraphics[width=0.45\textwidth]{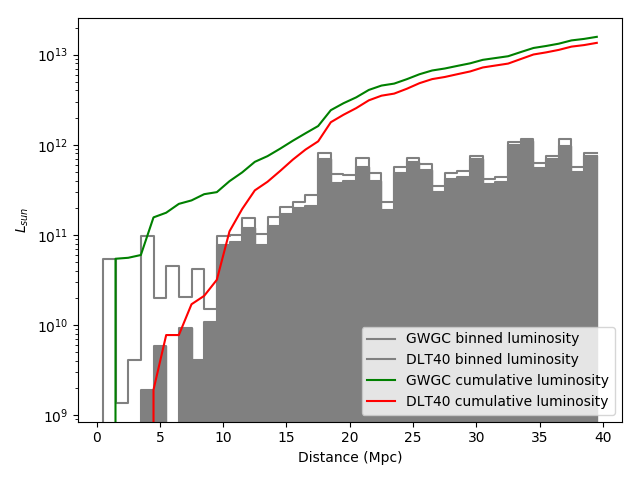}}
     \caption{Physical properties of the \ac{DLT40} galaxy sample compared with the GWGC \citep{GWGC}.}    
     \label{fig:dlt40cat}
   \end{figure*}

The \ac{DLT40} survey is designed as a high cadence supernova search. By targeting luminous galaxies within D$<$40 Mpc, we aim to find $\sim$10 supernovae per year within $\sim$1 day of explosion.
More details of the program are presented elsewhere \citep{Tartaglia18,dlt40rate} which we generally summarize here. 
During the Advanced Detector O2 run, \ac{DLT40} used a PROMPT 0.4m telescope \citep[PROMPT5;][]{PROMPT} which has a field of view (FoV) of 10$\times$10 arcmin$^2$ and is located at Cerro Tololo Inter-American Observatory (CTIO), part of the Skynet Robotic Telescope Network \footnote{https://skynet.unc.edu/}. In a typical night, $\sim$400-500 galaxies are observed.  The exposure time per field is 45 s, and the data is taken with no filter, yielding a typical limiting magnitude of $r$$\sim$19 mag. 
The \ac{DLT40} galaxy sample is selected from the Gravitational Wave Galaxy Catalogue \citep[GWGC;][]{GWGC}, with further cuts made on recessional velocity ($V$$<$ 3000 km s$^{-1}$ which corresponds to $D$$\lesssim$40 Mpc), declination (Dec$<$+20 deg), absolute magnitude ($M_{B}$$<$$-$18 mag), and Milky Way extinction ($A_V$$<$0.5 mag). We maintained our original \ac{DLT40} galaxy sample even after the Galaxy List for the Advanced Detector Era (GLADE\footnote{http://aquarius.elte.hu/glade}) was made available because the completeness of the two catalogues is not significantly different within 40 Mpc \citep{glade}.
The physical properties of the $\sim$2200 galaxies in the \ac{DLT40} sample are shown in Figure~\ref{fig:dlt40cat} in comparison to the whole GWGC sample within D$<$40 Mpc, selected without a luminosity cut. 

\begin{figure*}
\begin{center}
\includegraphics[width=\textwidth]{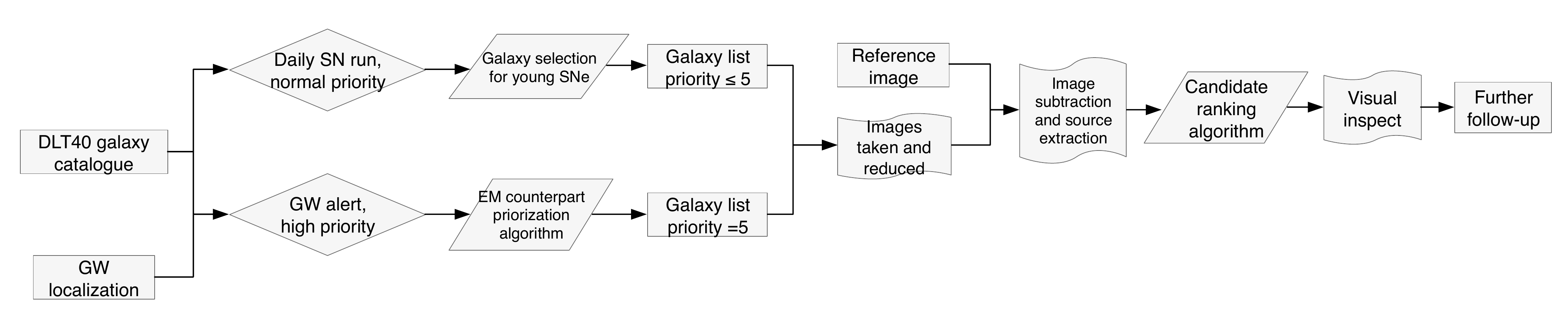}
\caption{Work flowchart for the DLT40 survey. \label{fig:fc}}
\end{center}
\end{figure*}

The DLT40 survey is fully robotic, and a flowchart of operations is shown in Figure~\ref{fig:fc}. A schedule is submitted automatically every afternoon before the Chilean sunset, and targets are given a priority between one and five. A score of five is the highest priority, and is reserved only for the most important targets such as the galaxies selected for \ac{GW} follow-up.  A score of four is assigned to galaxies that have been observed by the \ac{DLT40} survey over the last three days in order to maintain the program's cadence. A select few other galaxies are also given a native score of four -- for instance, if they are within $D$$<$11 Mpc, or if one PROMPT field of view can capture more than one \ac{DLT40} galaxy.  A score of three is assigned to other \ac{DLT40} galaxies not selected with higher priority, and which have $M_B$$<$$-$20 mag, while a score of two is assigned to those galaxies with $M_B$$<$$-$19 mag.  The remaining galaxies are given a score of one.  The Skynet scheduler observes targets from west to east within a priority category so that all of the priority five galaxies are observed first (if visible), followed by the priority four galaxies, and so on. Galaxy priorities can not be assigned in a `fine-grained' way beyond that described above, and so in this sense all of the galaxies targeted for the \ac{DLT40} \ac{GW} search were observed with an equally high priority, observed from west to east.

After an exposure is completed, the Skynet Robotic Telescope Network system automatically detrends the data (i.e. applies bias and flat field corrections), and determines an astrometric world coordinate system solution before the image is ingested by the \ac{DLT40} pipeline.  From there,  image subtraction is performed with respect to a high quality template image using the publicly available {\tt Hotpants} code \citep{hotpants}.
{\tt SExtractor}\footnote{http://www.astromatic.net/software/sextractor} \citep{sextractor} is then used to extract all sources in the difference images above a signal to noise threshold of five.
The difference image source catalog typically includes a large number of spurious objects due to stochastic processes, small misalignments between the images, improper flux scalings, imperfect PSF matching between the template and target image, and cosmic rays. In order to filter out spurious candidates, a scoring algorithm was developed based on catalog parameters returned by {\tt SExtractor}.
This approach still required visual screening of a significant number of candidates, most of which are rejected.  Recently we have implemented a machine learning algorithm for better filtering the spurious objects, which we discuss briefly along with our plans for the third observing run of the Advanced Detectors in Section~\ref{sec:o3}. In order to manage our real time data stream we use a {\tt MySQL\footnote{https://www.mysql.com/}} database and visually inspect SN candidates through web pages powered by the {\tt Flask\footnote{http://flask.pocoo.org/}} tool. 
After eyeballing, we secure immediate follow-up photometry or spectroscopy from collaborating facilities, most notably Las Cumbres Observatory, which itself is operated robotically \citep{Brown13}. 

The real time and quick response of the \ac{DLT40} SN search make it ideal for rapidly evolving transients including the \ac{EM} counterparts to \ac{GW} sources.  As such, \ac{DLT40} joined the global search effort during the Advanced Detector O2 run.  We discuss our \ac{GW} follow-up strategy in more detail next.

\section{GW follow-up strategy}
\label{sec:GWfollow}
The localization regions for \ac{GW} events are still formidably large (especially compared to the DLT40 FoV), being of order $\lesssim$1000 deg$^2$ for two detector events and $\sim$10-100 deg$^2$ for three \citep[e.g.][]{GWem}.  Two complementary \ac{EM} search strategies have been implemented by various groups thus far.  The first involves direct tiling of the \ac{GW} localization with wide-field telescopes in the search for counterparts \citep[e.g.][among others]{Smartt16,Kasliwal16,GWem,Soares17,Doctor18}, 
while the second relies on a targeted galaxy search focused on systems at a distance consistent with the \ac{GW} signal that are also within the localization region \citep[e.g.][]{Gehrels16}.  The \ac{DLT40} program followed the galaxy-targeted approach, although we relaxed any distance constraints whenever a \ac{GW} event was beyond the nominal $\sim$40 Mpc horizon of the main SN search, as we describe below.  The \ac{DLT40} PROMPT5 telescope's field of view (10'$\times$10') corresponds to 20 kpc (40 kpc) at a distances of 20 Mpc (40 Mpc), which encompasses $>$90\% of expected merger-host offsets \citep{Fong13}, and so each galaxy can be targeted with a single pointing.

\subsection{Galaxy prioritization}\label{sec:gal_prior}

The \ac{DLT40} software suite ingests the GCN alerts employed during O2 for disseminating \ac{GW} event information, and we download the {\tt HEALPIX} localization map with distance constraints \citep[see][for further information on the generation of these maps]{SingerPrice16,bayestar}.
From this GW-based data, we prioritize galaxies in the \ac{DLT40} catalog given the position, relative probability within the localization map, and the galaxy's inferred mass.  Again, we ignore any distance constraints from the GW signal in our galaxy prioritization algorithm (except for the nearby event GW170817). The target prioritization process is implemented as follows (see also Figure~\ref{fig:G275404}): 

\begin{figure*}
\begin{center}
\includegraphics[width=.6\textwidth]{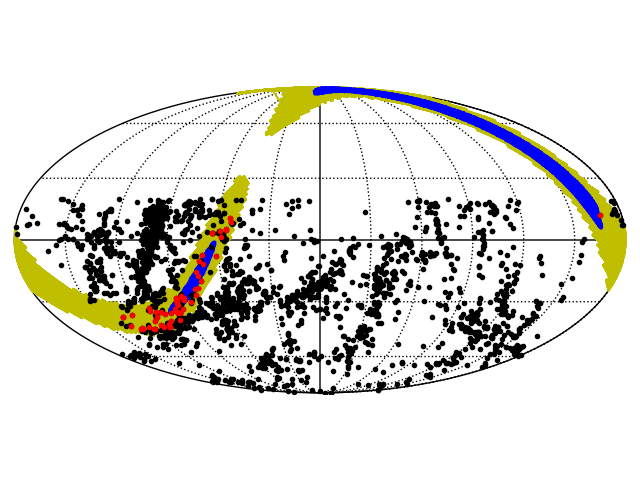}
\caption{Illustration of the  \ac{DLT40} galaxy ranking algorithm as applied to the LIGO trigger G275404. The full \ac{DLT40} galaxy sample is overlaid as black dots on the {\tt HEALPIX} \ac{GW} probability map; the blue and yellow region presents the 68\% and 95\% \ac{GW} confidence region, respectively.  After normalization, a galaxy's score is defined as the accumulated pixel values inside the galaxy area, convolved with the GW probability map. The top ranked galaxies are selected for immediate follow-up and monitoring as described in Section~\ref{sec:GWfollow}.  In the case of G275404, the red dots show the selected galaxies -- not all DLT40 galaxies within the LIGO 95\% confidence region were observed.  We observed up to a cumulative score of 0.99 (50 galaxies) in the initial phase of our follow-up (see Section~\ref{sec:G275404}).   }
\label{fig:G275404}
\end{center}
\end{figure*}

1. The \ac{DLT40} galaxy catalog is mapped with the {\tt HEALPIX} tool, weighted by galaxy luminosity (we assume that the mass distribution follows the $B$-band luminosity), and smoothed with a Gaussian corresponding to each galaxy's reported radius.  This yields a luminosity distribution map of \ac{DLT40} galaxies, $S_{lum}$, which we normalize ($N_{lum}$):
\begin{equation}
N_{lum} = \frac{S_{lum}}{\sum{S_{lum}}}
\end{equation}

2. The \ac{GW} probability map from LVC is already normalized as $S_{gw}$.

3. We then take the product of the \ac{DLT40} galaxy map and the \ac{GW} probability map,
\begin{equation}
C = S_{gw} \times N_{lum}
\end{equation}

4. For each \ac{DLT40} galaxy $i$, we sum over all pixels in $C$ inside the galaxy radius, which yields our prioritization, 
\begin{equation}
s_{i} = \sum_{j}{C_{ij}}
\end{equation}
where $j$ is the {\tt HEALPIX} pixel value inside the specific radius.  

\begin{figure*}
\begin{center}
\includegraphics[width=.8\textwidth]{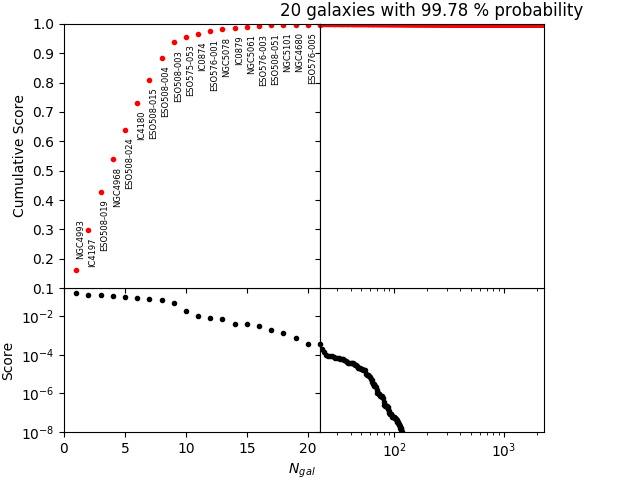}
\caption{Cumulative score for DLT40 galaxies as prioritized for the NS-NS binary GW170817, using the algorithm described in Section~\ref{sec:GWfollow}. 
As shown, NGC~4993, the host of the kilonova DLT17ck/AT2017gfo, is ranked first.  The twenty selected DLT40 galaxies cover 99.78\% of the cumulative probability. More detailed information for each galaxy is presented in Table~\ref{GW170817}. 
}
\label{fig:rank2}
\end{center}
\end{figure*}

After the galaxy priorization procedure, a cut on the number of monitored galaxies was made based on the level of interest in the GW trigger and practical observing considerations.  For some GW sources with very large localization regions, we would also make cuts based directly on these maps -- for instance, only selecting galaxy targets within the 80\% credible region; see Section~\ref{sec:O2} for details on individual sources.  We would select galaxies by cutting the ranked list at the top 50\% - 99\% of the normalized cumulative score, $s$, which would then be further reduced since some galaxies in the list would be unobservable (i.e. below the horizon).  For instance, if the  \ac{GW} trigger indicated the source was a local real (with a very low FAR) CBC, we would select galaxies up to a cumulative score of 99\%, as shown in Figure \ref{fig:rank2}.  For those GW triggers that DLT40 followed up in O2, we observed between 18 and 114 galaxies per event, which corresponded to cumulative scores between 0.1 and 0.99.
All the selected galaxies were set to priority 5 in the DLT40 scheduler.

\begin{figure*}
\begin{center}
\includegraphics[height=.8\textwidth]{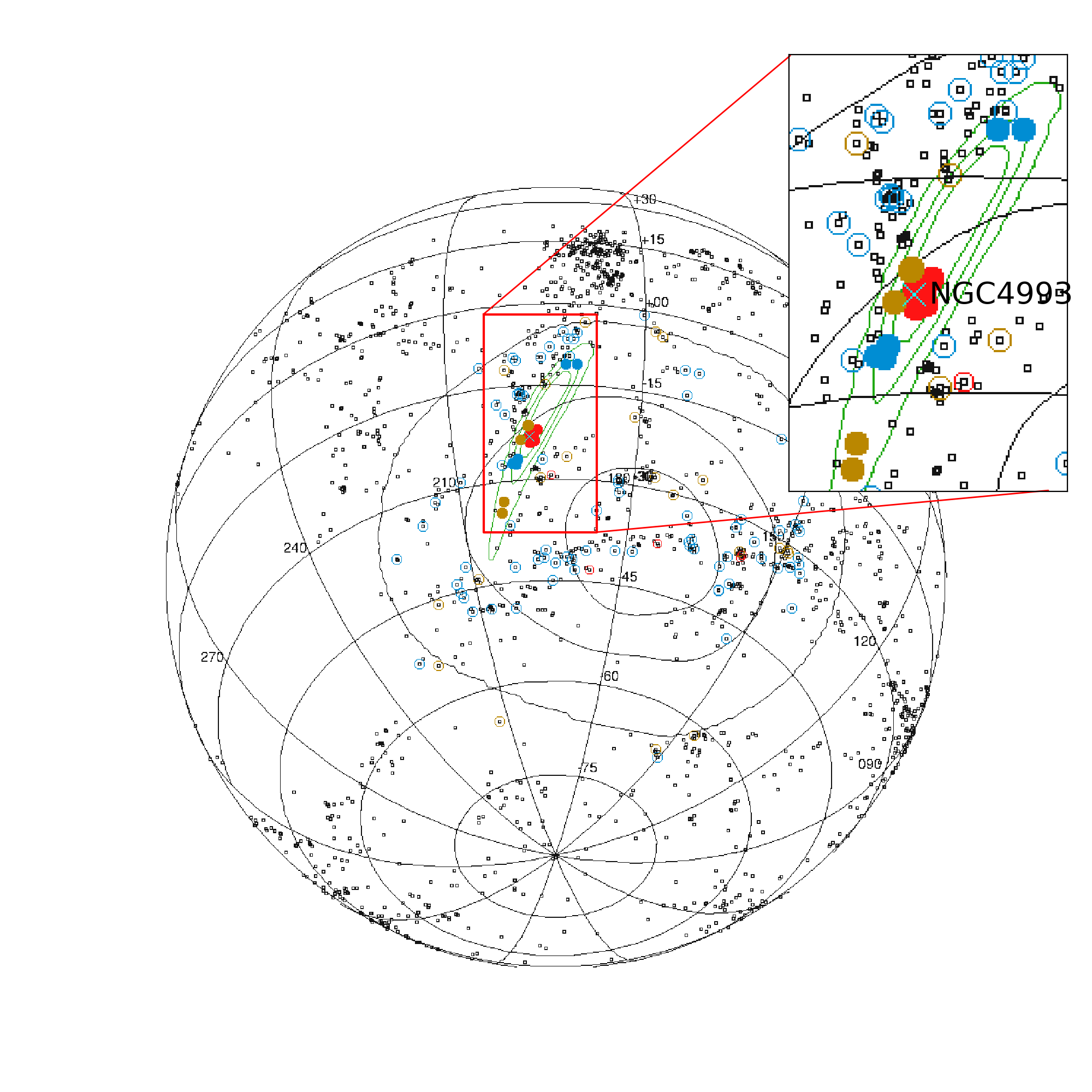}
\caption{Localization region (contours) and the matched galaxies (circles) for GW170817/G298048 (solid circles) and GRB170817a (hollow circles). The contours indicate 50\%, 90\%, and 99\% confidence bounds while the \ac{GW} trigger is shown in green and the GRB trigger is shown in black. The colors of the circles denote the priority of the galaxies (high priority in red, normal priority in yellow and low priority in blue). All the \ac{DLT40} galaxy samples are shown as black dots. After ranking, we decided to follow all 20 \ac{GW} galaxies (9 high + 5 normal + 6 low) and the top 31 GRB galaxies (5 high + 26 normal).}
\label{fig:rank}
\end{center}
\end{figure*}

As evidence of the efficacy of our galaxy prioritization algorithm, the highest priority galaxy for the binary neutron star event GW170817 was NGC~4993, the true host galaxy of the counterpart (see Section~\ref{gw170817} and Table~\ref{GW170817} for further details).  We discuss each individual GW event and our detailed follow-up from O2 further in Section~\ref{sec:O2}.

In principle, we can adjust our galaxy prioritization to account for the GW-inferred distance (if given) and other detection efficiencies \citep[similar to][]{lcogt}.  For the LVC O2 run, we did not include the GW-inferred distance into our galaxy targeting algorithm (which were only available for CBC triggers), but will do so for the DLT40 O3 campaign (Section~\ref{sec:o3}).

\subsection{Monitoring timescale}

\begin{figure*}
\begin{center}
\includegraphics[width=.8\textwidth]{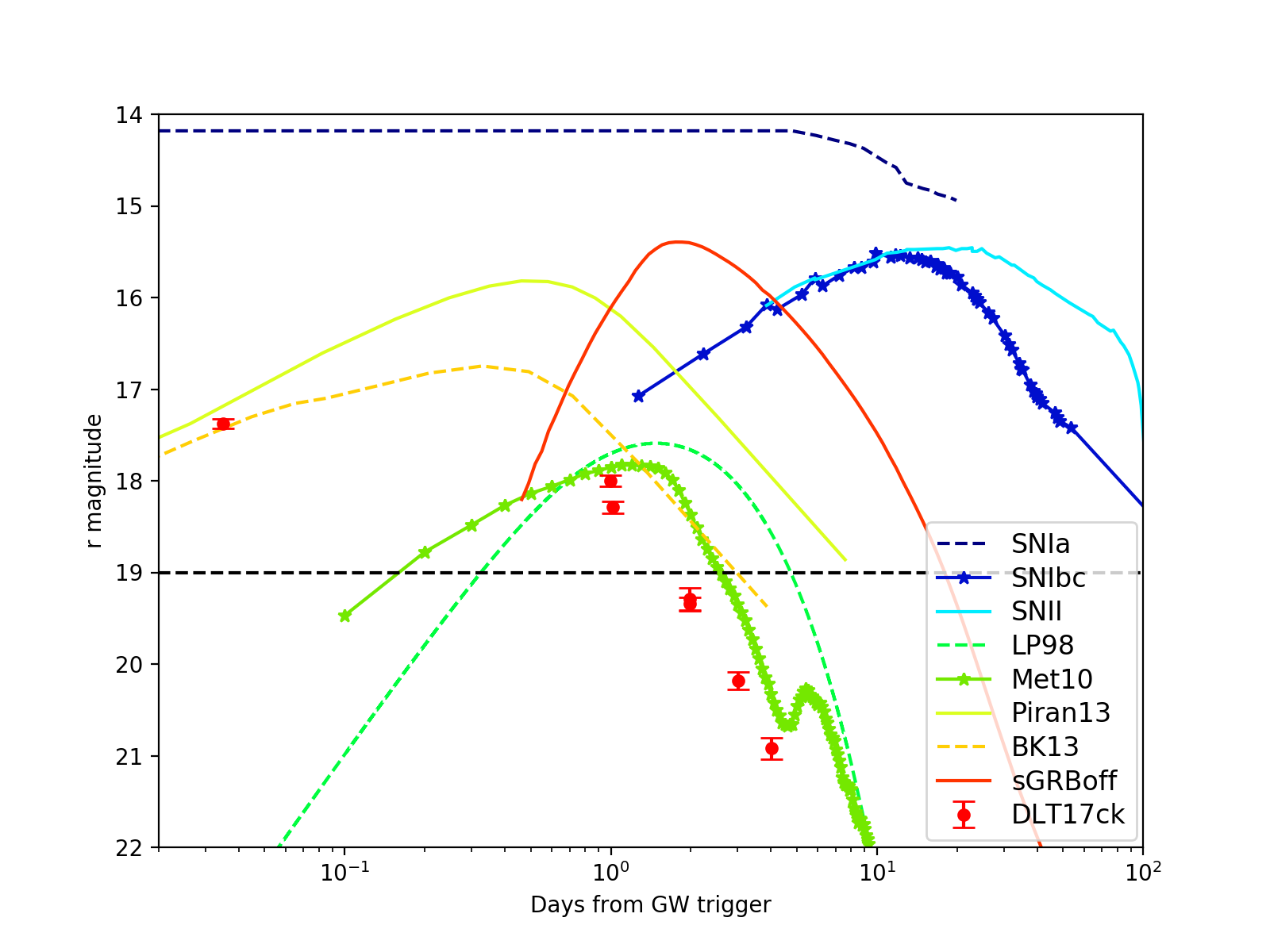}
\caption{Several possible EM emission models of \ac{GW} sources, scaled to a distance of 40 Mpc, are plotted against the 6 epochs of DLT17ck observations (red dots), as measured by the DLT40 survey \citep{dlt17ck}. Among these optical emission models, there are four kilonova models: LP98 \citep{LP}, assuming a blackbody emission for an ejecta mass $10^{-2} M_{\sun}$, and outflow speed of $v = 0.1c$; Met10 \citep{Metzger10}, assuming a radioactively powered emission with an ejecta mass $10^{-2} M_{\sun}$, and outflow speed of $v = 0.1c$, as well as iron-like opacities; BK13 \citep{B&K}, assuming an ejected mass of $10^{-3} M_{\sun}$ and velocity of $0.1 c$, along with lanthanide opacities; Piran13 \citep{piran}, assuming a BH-NS merger with $NS=1.4 M_{\sun}$, $BH=10 M_{\sun}$.  We also include one sGRB off axis model: sGRBoff \citep{sGRBoff1}, a simulated off-axis afterglow light curve assuming a short GRB with ejecta energy of $E_{jet} = 10^{50} erg$, interstellar matter density of $n =10^{-3} cm^{-3}$, jet half-opening angle of $\theta_{jet} = 0.2$ radians and an observed viewing angle of $\theta_{obs} = 0.2$ radians. For comparison, we also show three SN light curves, including SN 2012cg \citep[a type Ia SN;][]{12cg}, SN2002ap (a type Ib/c SN) and SN SN2013ej \citep[a type II SN;][]{13ej}.}
\label{fig:model}
\end{center}
\end{figure*}

The duration that we monitor GW-related galaxies depends on the type of trigger.  For burst-type \ac{GW} events that are possibly related to core-collapse supernovae (with an optical time scale of $\sim$10-100 days), we adopt a monitoring period of three weeks.  For binary neutron star-type mergers, an r-process kilonova \citep{LP,Metzger10,Kasen13} and/or  short gamma-ray burst (sGRB) afterglow emission is expected. Figure \ref{fig:model} shows different light curve models for kilonovae and sGRB afterglow emission, scaled to a distance of 40 Mpc.  As shown, DLT40 could detect transients predicted by most of the models out to this distance (the limiting magnitude of \ac{DLT40} is $r$$\sim$19 mag, as measured by artificial star experiments, described in \citealt{dlt40rate}).  We thus conservatively monitored compact binary events for $\sim$2 weeks, well beyond the time we would generally expect to have a detection.

\subsection{Summary}
This galaxy prioritization process produces a list of galaxies that are sent automatically to the \ac{DLT40} scheduler.  As discussed, these galaxies are observed at the highest priority, going from west to east on the sky, and are immediately ingested into our transient detection pipeline.  All transient candidates were scored with our image feature-based algorithm and forwarded within minutes of data taking for visual inspection.  If an interesting transient was found, a GCN was issued as soon as possible to facilitate follow-up observations; if no viable candidates were found, a summary GCN was usually posted in the following days to weeks.

\section{Search results in O2}
\label{sec:O2}
\begin{figure*}
\begin{center}
\includegraphics[width=.8\textwidth]{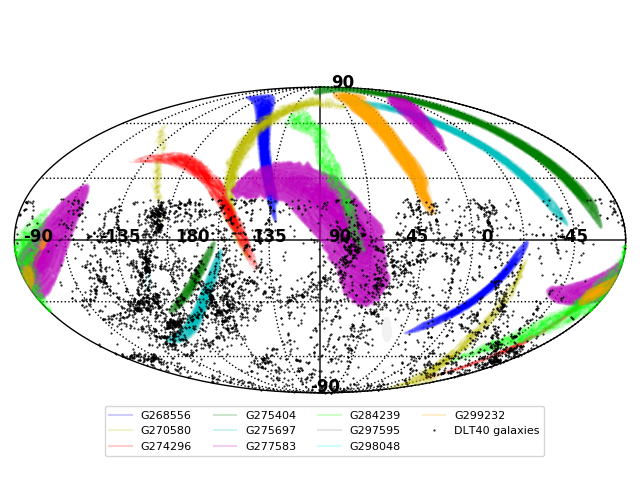}
\caption{The \ac{DLT40} galaxy sample super-imposed on the 68\% confidence contours of the ten LVC triggers followed by \ac{DLT40} during O2. See Section~\ref{sec:O2} and Table~\ref{tabsummary} for further details. }
\label{fig:maps}
\end{center}
\end{figure*}

The LIGO O2 observing run ran from 2016 November 30 to 2017 August 25, with Virgo joining the network of GW detectors in early August, 2017. 
Fourteen GW alert triggers were issued (six of which turned out to be real events) by the LVC for follow-up to the EM community \citep{lvc_19},  among which we followed the ten triggers indicated in Table~\ref{tabsummary} with their localization shown in Figure \ref{fig:maps}.  Three of these ten events that DLT40 followed up were ultimately found to be true GW events after further GW analysis (these are also marked in Table~\ref{tabsummary}).  
%Notably, in LVC O2 run we decided to follow all the \ac{GW} triggers (if possible), to probe the uncertain emission from BBH.
%if possible, since till the moment, whether there're any \ac{EM} radiations from \ac{BBH} is still undecided \citep{bbhem}.

DLT40 pursued BBH events without prejudice during O2, even though these events may not have associated EM emission. 
GW horizon distances of \ac{BBH} events are typically relatively large compared to the nominal reach of the \ac{DLT40} survey (focused on galaxies within 40 Mpc).
For instance, the initial luminosity distance of G275697 estimated by the LVC has a mean value of 181 Mpc and standard deviation 55 Mpc \citep[see Table~\ref{tabsummary};][]{lvcgcn275697a}. The cumulative probability within a 40 Mpc volume is only 1\%. However, we activated our follow-up search for GW events outside the nominal  DLT40 distance horizon for the following reasons:

\begin{itemize}

\item As mentioned, during the LVC O2 run, the distance estimates of CBC are generally available, but are sometimes updated later or not reported in early circulars.

%\item In case of new physics that results in poor distance estimates from CBC waveforms.

\item DLT40 is not dedicated to GW follow-up, so there is no conflict with our daily SN search -- it does not harm us to search even if the probability of success is low. 

\item The search could be implemented as a test run in preparation for events that DLT40 is more likely to find counterparts for \citep[e.g. GW170817;][]{dlt17ck}.

\end{itemize}

\noindent For such BBH events with estimated GW distance beyond the reach of DLT40, we activated only the top 50\% of galaxies scored by our ranking algorithm (see Section~\ref{sec:GWfollow}).
 
Here we present our observations for the ten DLT40-followed \ac{GW} triggers, focusing first on the two events where our team identified optical transients.  Unless otherwise noted, all dates listed are UTC.  Detailed observational logs for each event are presented as a series of tables in the Appendix.

\startlongtable
\begin{deluxetable*}{cccccccccc}
\tablecaption{Summary table of DLT40 follow-up of LVC O2 triggers. GW alert properties include the source type, the false alarm rate (FAR), the 90\% confidence level sky localization area, and the luminosity distance. In the second column block, we report DLT40 observations, including the number of galaxies observed, their observation window, possible transients detected by DLT40 in that dataset, and the reference GCNs. It should be mentioned that the GW estimates shown here are taken from the LVC preliminary and updated GCNs, the final GW estimation is shown in \cite{lvc_19}. \label{tabsummary}}
\tablehead{
& \multicolumn{5}{c}{LVC GW estimations} & \multicolumn{4}{c}{DLT40 EM observations} \\
GW/GRB  & Source type & FAR$^{a}$ & Localization$^{b}$ & $D_{lum}$ & ref$^{c}$ & $N_{gal}$ & $Obs\_window$ & Detection$^{d}$  & REF$^{e}$ \\
trigger  & & (yr$^{-1}$) & (deg$^{2}$) & (Mpc)&   & &  (JD) & or $r_{lim}$ & 
%\colhead{GW/GRB}  & \colhead{Source type} & \colhead{FAR\tablenotemark{a}} & \colhead{Localization\tablenotemark{b}} & \colhead{$D_{lum}$} & \colhead{ref$^{c}$} & \colhead{$N_{gal}$} & \colhead{$Obs\_window$} & \colhead{Detection$^{d}$} & \colhead{REF$^{e}$} \\
%\colhead{trigger} & \colhead{} & \colhead{(yr$^{-1}$)} & \colhead{(deg$^{2}$)} & \colhead{(Mpc)} & \colhead{} & \colhead{} & \colhead{(JD)} & \colhead{or $r_{lim}$} & \colhead{}
}
%\colnumbers
\startdata
GW170104  $^{f}$ & CBC & 2 & 1600  & \nodata & a & 18 & 2457759- & \nodata & \nodata \\
(G268556)  &  &  &   &  &  &  & 2457766 &  &  \\
%\hline
G270580 $^{f}$ & Burst & 5 & 3100 & \nodata & b c & 25 &2457774-  & 19.2 & A \\
&  &  &  &  &  &  &2457796  &  &  \\
%\hline
G274296 & Burst & 6 & 2100 & \nodata & d e & 25 &2457802  & 18.5 & B \\
 &  &  &  &  &  &  &2457825  &  &  \\
%\hline
G275404 & CBC &6 & 2100 & 280$\pm$80 & f & 50 & 2457815-  & DLT17u  & C\\
 &  & &  & &  &  & 2457821  &  SN2017cbv & \\
        & & & 17000 &  \nodata & g & 84 & 2457821-  & (SN Ia) & \\
       & & &  &   &  &  & 2457825  &  & \\
%\hline	
G275697 & CBC & 6 & 1800 & 181$\pm$55 & h & 59 & 2457812- & 19.0 & D\\
 &  &  &  &  &  &  & 2457820 &  & \\
        & & & 3890  &  193$\pm$61  & i & 114 &2457820- & &\\
        & & &  &    & &  &2457825 & &\\
%\hline
G277583 & Burst & 3 & 12140 & \nodata & j & 55 &2457826-& 19.5 & E\\
 &  & &  & & &  &2457847&  & \\
%\hline
G284239 & Burst & 4 & 3593 & \nodata & k & 58 &2457877-& 19.1 & F\\
 &  &  &  &  &  &  &2457892&  & \\
%\hline
GW170608  & CBC & 2.6 & 860 & 320 $\pm$ 100 & l & \nodata & \nodata & \nodata & \nodata \\
(G288732) &  &  &  &  &  &  & &  &  \\
%\hline
GW170809  & CBC & 0.25 & 1155 & 1000 & m & \nodata & \nodata & \nodata & \nodata \\
 (G296853) &  &  &  &  &  & &  & &  \\
%\hline
GW170814 & CBC & 1/82800 & 97 & 550$\pm$130 & n & 24 & 2457980- & 19.0 & G\\
 (G297595) &  &  &  &  &  &  & 2457982 &  & \\
%\hline
GW170817  & CBC & 1/9100 & 33.6 & 40 $\pm$ 8  & o p & 20 & 2457983-  & DLT17ck  & H I J\\
 (G298048) &  &  &  &   &  &  & 2457985  &  AT2017fgo (KN) & \\
GRB170817a &&&&& q & 31 &&& \\
%\hline
G298389 & Burst & 6 & 799  & \nodata & r & \nodata & \nodata & \nodata & \nodata \\
%\hline
GW170823  & CBC & 1/1800 & 2145 & 1387 $\pm$ 414 & s & \nodata & \nodata & \nodata & \nodata \\
 (G298936) &  &  &  &  &  &  & &  &  \\
%\hline
G299232 & CBC & 5.3 & 2040 & 339 $\pm$ 110 & t & 20 &2457991- & 19 & \nodata \\
 &  &  &  &  &  &  &2458005 &  &  \\
\enddata
\tablenotetext{a}{The ``FAR'' column is the estimated false alarm rate per year, while the LVC threshold is one per month or 12 per year. If the FAR is less than 12, an alert would be issued.}
\tablenotetext{b}{The ``localization'' column is the   uncertainty area of GW triggers in 90\% confidence level.}
\tablenotetext{c}{a.\cite{lvcgcn268556}; b.\cite{lvcgcn270580a}; c.\cite{lvcgcn270580b}; d.\cite{lvcgcn274296a}; e.\cite{lvcgcn274296b}; f.\cite{lvcgcn275404a}; g.\cite{lvcgcn275404b}; h.\cite{lvcgcn275697a}; i.\cite{lvcgcn275697b}; j.\cite{lvcgcn277583}; k.\cite{lvcgcn284239}; l.\cite{lvcgcn288732}; m.\cite{lvcgcn296853}; n.\cite{lvcgcn297595}; o.\cite{lvcgcn298048a}; p.\cite{lvcgcn298048b}; q.\cite{lvcgcn_fermi_gwfollow}; r.\cite{lvcgcn298389}; s.\cite{lvcgcn298936}; t.\cite{lvcgcn299232}.}
\tablenotetext{d}{We report the typical DLT40 limiting magnitude (r band) for the observation set in the ``Detection'' column.}
\tablenotetext{e}{A.\cite{gcn270580}; B.\cite{gcn274296}; C.\cite{gcn275404}; D.\cite{gcn275697}; E.\cite{gcn277583}; F.\cite{gcn284239}; G.\cite{gcn297595}; H.\cite{gcn170817a}; I.\cite{gcn170817b}; J.\cite{gcn170817c}.}
\tablenotetext{f}{For GW170104 (G268556) and G270580 we only observed DLT40 galaxies out to 20 Mpc within the LVC localization as a test run.}
\end{deluxetable*}

\subsection{G275404}
\label{sec:G275404}
G275404 was identified as a marginal GW candidate by the two LIGO interferometers, Hanford (H1) and Livingston (L1) using the pyCBC analysis \citep{pycbc} on 2017-02-25, 18:20:21.374 UTC (GPS time: 1172082639.374). The false-alarm rate of $1.89\times 10^{-7} Hz$ corresponds to $\sim$1 in 0.17 years. Following the initially-released bayestar localization map \citep{bayestar}, the 50\% (90\%) credible region spanned about 460 (2100) deg$^{2}$.
DLT40 responded rapidly (within an hour) to select 50 galaxies within the LVC error region, which were observed from 2017-02-26 to 2017-03-09.
At 2017-03-08 22:30:50 UTC (11 days after the initial trigger), the LALInference localization map \citep{lal} was issued by the LVC with a 50\% (90\%) credible region which increased to about 2000 (17000) deg$^{2}$. 
Meanwhile, the LVC announced that the binary component masses were estimated to be consistent with a \ac{BNS} or NS-BH binary \citep{lvcgcn275404b}. 
Starting on 2017-03-09 (until 2017-03-12) we re-prioritized our targeted galaxy list, and observed 84 galaxies using the updated \ac{GW} localization map. 
The first 50 galaxies represent 3.9\% of all galaxies in the Glade catalogue within 40 Mpc and contain 12.7\% of all B band luminosity of those galaxies; the latter 84 galaxies represent 1.5\% of all galaxies and contains 9.7\% of all B band luminosity. 
The typical \ac{DLT40} limiting magnitude for these observations was 19.2 mag (open filter scaled to r band). 

We found one SN Ia, SN2017cbv/DLT17u, in NGC5643 (at a distance of $D$$\approx$12.3 Mpc \cite{Sand18} in the GW follow-up observations taken on 2017-03-08. 
Our follow-up observations of SN2017cbv/DLT17u with Las Cumbres Observatory telescopes show that SN2017cbv/DLT17u reached its maximum brightness (B=11.79 mag) 17.7 days after discovery \citep{Hosseinzadeh17}. 
Given the typical rise time of SNe Ia \citep[$18.98 \pm 0.54$ days; ][]{sn1a_lc}, we deduced that SN2017cbv was discovered very close to the explosion epoch and we can thus exclude SN2017cbv from being related to the \ac{GW} event,  which occurred $\sim$2 weeks prior.
Ultimately, further LVC analysis indicated that G275404 was not a trigger of interest \citep{lvc_19}, confirming that SN~2017cbv was an unrelated transient.

\subsection{GW170817/G298048}
\label{gw170817}
GW170817 \citep{GW170817,GW170817MMA} was identified by the LIGO H1 detector at August 17, 2017 12:41:04 UTC (GPS time: 1187008882.4457), as a likely \ac{BNS} merger event. 
The false alarm rate was $3.478\times10^{-12}$ Hz, equivalent to $\sim$1 false alarm in 9100 years.
In addition, the GW candidate was found in coincidence with the \textit{Fermi}/GBM trigger 524666471/170817529 \citep[GRB170817a; ][]{GRB170817a},  registered about 2 seconds later on Aug 17, 2017 12:41:06 UTC (GPS time: 1187008884.47).
With further constraints from Virgo, the LVC joint sky area was reduced to 8.6 (33.6) deg$^{2}$ for the 50\% (90\%) credible regions, making it the best constrained event of the  full O2 run.
%Actually, all the inferences at the moment indicate that G298048 was a very interesting GW trigger.
As shown in Figure \ref{fig:rank}, 
inside the 90\% confidence region of GW170817 (33.6 $deg^{2}$), there were only 23 galaxies based on our \ac{DLT40} galaxy selection, and
we decided to observe 20 of them (with 99\% galaxy priorization cut), together with the 31 most luminous galaxies in the GRB error-box region (Fermi GBM trigger). 
About $\sim$11 hours after the announcement of the \ac{GW} trigger, at the beginning of the Chilean night, \ac{DLT40} reported the detection of DLT17ck at RA=13:09:48.09 and DEC=-23:22:53.4.6, 5.37W, 8.60S arcsec from the center of NGC 4993 \citep{dlt17ck,gcn170817a,gcn170817b,gcn170817c}.
Within one hour, we were one of the six optical groups which independently detected this optical transient, also named AT 2017 gfo and SSS17a \citep{GW170817MMA,sss17a,dlt17ck,vista,master,decam,lcogtkn}.   It was soon established that 
DLT17ck was a GW-associated kilonova \citep{LP,grawita,Villar17,Evans17,Metzger17,Kasen17}.
No other transients were found in the other surveyed galaxies that DLT40 followed in the GW170817 localization region. 
The DLT40 follow-up data obtained for DLT17ck are described in \cite{dlt17ck}.
\cite{dlt40rate} further used the observed light curve of DLT17ck to constrain the rate of \ac{BNS} mergers to be less than 0.50 SNuB\footnote{$\rm{SNuB}=1\,\rm{SN}$ per $100\,\rm{yr}$ per $10^{10}L_{B_{\odot}}$} and we concluded that \ac{DLT40} would need to operate for $\sim 18.4$ years in order to discover a kilonova without a \ac{GW} trigger.

%\begin{figure*}
%\begin{center}
%\includegraphics[width=.8\textwidth, angle=90]{lightcurves}
%\caption{Top: SN and KN identified in our survey during O2. Black: SN Ia, 2017cbv/DLT17u discovered after G275404. Red: KN, AT17fgo/DLT17ck discovered after GW170817/G298048. In all images the showed field sizes are $10\times10$ arcsec, North is up and East to the left. The cross represent the position identified by our pipelines. Bottom: The light curves of DLT17u and DLT17ck while the data before the \ac{GW} discovery(dashed line) represents the magnitude upper limit of the the host galaxy.}
%\label{fig:dlt17ck}
%\end{center}
%\end{figure*}

\subsection{Other \ac{GW} triggers}

GW170104/G268556 \citep{GW170104,lvcgcn268556} was identified by L1 and H1 at 2017-01-04 10:11:58.599 UTC (GPS time: 1167559936.599). GW170104 was a \ac{BBH} event with a low false-alarm rate, 6.1$\times$10$^{-8}$ Hz (about one in 6 months). 
This \ac{GW} event was the first identified LVC trigger in the O2 run, and the first \ac{GW} event followed by DLT40. The initial announcement of GW170104/G268556 did not include a distance estimate.  
For this first event, we decided to test our galaxy ranking algorithm and other infrastructure as a dry run, and not report our findings via a GCN.
We monitored 18 galaxies within 20 Mpc for $\sim$1 week within the 90\% GW localization region, and no optical counterpart candidates were found.

G270580 \citep{lvcgcn270580a,lvcgcn270580b} was identified by L1 and H1 at 2017-01-20 12:30:59.350 UTC (GPS time: 1168950677.350). The false alarm rate was 1.6$\times$10$^{-7}$ Hz (about one in 2.4 months). The 50\% credible region spanned about 600 deg$^2$ and the 90\% region about 3100 deg$^2$. 
We observed 25 galaxies from the \ac{DLT40} catalogue within the 99\% credible region, again within a distance of 20 Mpc, from 2017-01-23 onwards, for three weeks. No interesting transients were identified down to a typical limiting magnitude of 19.2 mag in r band \citep{gcn270580}. This GW event was ultimately retracted by the LVC.

G274296 \citep{lvcgcn274296a,lvcgcn274296b} was identified as a burst candidate by H1 and L1 at 2017-02-17 06:05:55.050 UTC (GPS time:1171346771.050), with false alarm rate 1.7$\times$10$^{-7}$ Hz, or about one in 2 months. 
We observed 25 galaxies from our \ac{DLT40} galaxy catalogue (out to the full DLT40 D=40 Mpc search volume) within the 80.0\% credible region of the trigger localization region. 
We began observing these  galaxies on 2017-02-17 and monitored them for 3 weeks after the \ac{GW} trigger. No interesting transients were identified down to an average limiting magnitude of 18.5 \citep{gcn274296}.  The LVC ultimately classified this GW trigger as of no further interest.

G275697 \citep{lvcgcn275697a,lvcgcn275697b} was identified as a marginal candidate by L1 and H1 at 2017-02-27 18:57:31.375 UTC (GPS time: 1172257069.375), with a false alarm rate of 1.43$\times$10$^{-7}$ Hz or about one in 2 months. Based on a preliminary analysis, the LVC reported that the less massive component in the binary had a mass $<$3 $M_{sun}$ and that there was a 100\% chance that the system ejected enough neutron-rich material to power an electromagnetic transient.
The 50\% credible region spanned about 480 deg$^2$ and the 90\% region about 1800 deg$^2$. The luminosity distance was estimated to be $181 \pm 55$ Mpc. 
We observed 59 galaxies from  2017-02-27 to 2017-03-07. After an update of the LVC localization map, we updated the galaxy sample and observed 114 galaxies from 2017-03-07 to 2017-03-12. 
No interesting transients were identified down to a limiting  magnitude of $\approx$19.0 \citep{gcn275697}.
The final analysis of the LVC indicated that G275697 was not a trigger of interest.

G277583 \citep{lvcgcn277583} was identified as a burst candidate by L1 and H1 at 2017-03-13 22:40:09.593 UTC (GPS time:1173480027.593). Its false alarm rate was 8.4$\times$10$^{-8}$ Hz (one in 4 months). We observed 55 galaxies  within the 80.0\% credible region, beginning on 2017-03-13 and continuing for 2 weeks after the \ac{GW} trigger. No interesting transients were identified down to a limiting magnitude of 19.5 mag \citep{gcn277583}.  The final analysis of the LVC indicated that G277583 was not a trigger of interest.

G284239 \citep{lvcgcn284239} was identified by L1 and H1 at 2017-05-02 22:26:07.910 UTC (GPS time: 1177799185.910). G284239 was a low-significance short-duration burst candidate, whose false alarm rate was 1.26$\times$10$^{-7}$ Hz (4 per year). The 50\% confidence region covered 1029 $deg^2$ and the 90\% confidence region covered 3593 $deg^2$. We observed 58 galaxies within the 95\% credible region of the localization for a duration of two weeks starting on 2017-05-02 \citep{gcn284239}.  No interesting transients were identified down to a limiting magnitude of $r$$\sim$19 mag.  The LVC determined that this event was of no further interest.

G297595/GW170814 \citep{lvcgcn297595} was the first \ac{GW} event detected by both the two LIGO detectors (H1, L1) and the Virgo (V1) detector \citep{GW170814} at 2017-08-14 10:30:43 UTC (GPS time: 1186741861.5268). 
The Virgo detection helped to decrease the 50\% (90\%) localization region from 333 (1158) deg$^{2}$ to 22 (97) deg$^{2}$. GW170814 had a very low false alarm rate of $3.83\times10^{-13}$ Hz, equivalent to $\sim$ 1 per 82800 years. The LVC reported that the event was a \ac{BBH} merger at $\sim 550 \pm130$ Mpc. 
Despite the lack of an expected optical counterpart and the large distance, we triggered follow up partly because of the small localization region.
We monitored 24 galaxies within the LVC error region with an average limiting magnitude of 19.0 mag \citep{gcn297595}. No obvious optical counterparts were detected. 
All selected galaxies from this trigger were reset to normal priority on 2017 August 17 in order to aggressively pursue the next trigger, GW170817.

G299232 \citep{lvcgcn299232} was identified by L1 and H1 at 2017-08-25 13:13:31 UTC (GPS time: 1187702035.9831). G299232 was a low-significance candidate with a false alarm rate of  1.68$\times$10$^{-7}$ Hz (about 5.3 per year). The 50\% credible region spanned about 450 deg$^2$ and the 90\% region about 2040 deg$^2$.  We selected and observed 41 galaxies within 95.0\% of the trigger localization region from 2017-08-25 to 2017-09-08. No obvious transient was found, and the LVC concluded that this event was of no further interest.

%\begin{figure*}
%\begin{center}
%\includegraphics[width=.8\textwidth]{lim_dist.png}
%\caption{Limiting distance for different EM counterpart models, assuming a limiting magnitude of $r$=19 mag for \ac{DLT40} observations. %obtained by artificial star experiments. With the assumption that the limiting magnitude for \ac{DLT40} measurements is r=19 mag, we scale modeling light curves into different distances while the limiting distance is defined as the distance where the specific model cannot reach the limiting magnitude in any epochs. 
%The models are described and shown as in Figure \ref{fig:model}. We didn't include SNIa model here, since its limiting distance is always above 200 Mpc even after 100 days.}
%\label{fig:limdist1}
%\end{center}
%\end{figure*}

\begin{figure*}
\begin{center}
\includegraphics[width=.8\textwidth]{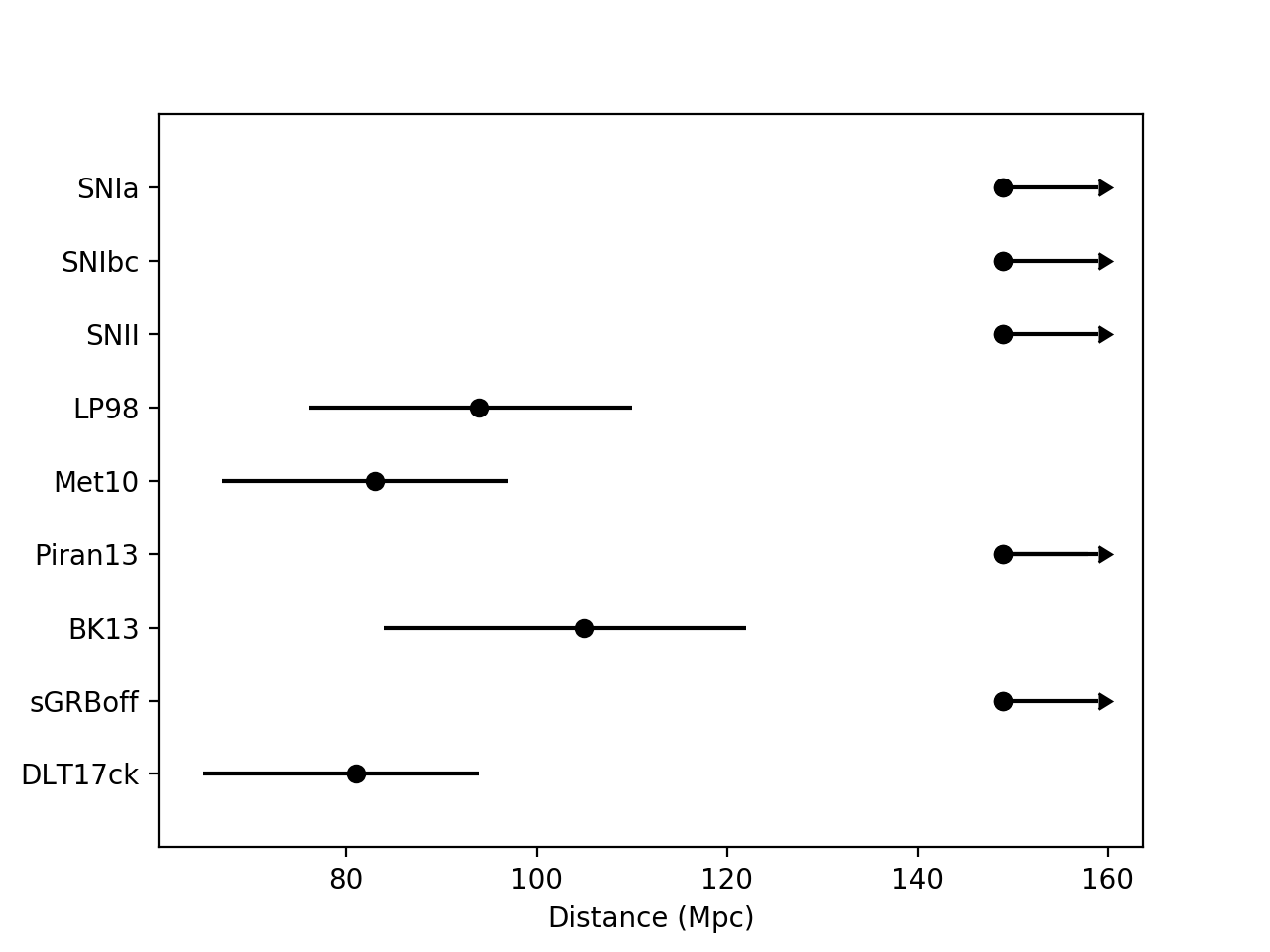}
 \caption{The DLT40 limiting distance for different BNS EM emission models (details described in Figure \ref{fig:model}), and the observational data of DLT17ck, are estimated.
 The computational recipe is described as followed:
 1. detection efficiency (DE) for single epoch/point: by setting up a series of artificial star experiments, we obtain the DE plots, which show the relations between the DLT40 detection efficiency, $DE$, with  magnitude, for a specific field \citep{dlt40rate};
 2. DE for epochs/sampling light curve: for a specific light curve (EM models or observations in absolute magnitudes), by executing step 1 for a sample of epochs \textsuperscript{a}, we obtain $DE_j$ for each epochs, while the overall DE is derived as $DE = 1-\Sigma_j(1-DE_j)$;
 3. limiting distance estimation: we scaled each light curve to a distance of $D_i$, and obtain $DE_{D=D_i}$ at different distances. The limiting distance is then defined as the distance where $DE_{D=D_i}$ is $50\%$.}
    \small\textsuperscript{a} We decide the sample points by comparing the burst time of GW170817, with our DLT40 observing epochs towards NGC4993.
\label{fig:limdist2}
\end{center}
\end{figure*}

\section{Future Prospects} 
\label{sec:future}
In this section, we describe the DLT40 project's follow-up plans for the upcoming LVC O3 run based on some simple simulations of GW events, and the performance of DLT40. %After these plans have been applied, we describe the possible scenarios that would be happened in O3, and our possible response.

\subsection{DLT40 EM Counterpart Potential}
\label{sec:up}

To evaluate DLT40's potential impact for O3, we made use of the artificial star tests described in \citet{dlt40rate}, which provide point source detection efficiencies as a function of image zeropoint and seeing.  Using these artificial star tests in conjunction with the real light curve of AT2017gfo/DLT17ck and the EM counterpart models shown in Figure~\ref{fig:model}, we calculate the limiting distance out to which the DLT40 program can detect each of these transients, as shown in Figure~\ref{fig:limdist2}.  Kilonovae with light curves similar to AT2017gfo/DLT17ck can be detected out to $D$$\approx$80 Mpc, while other kilonova light curve models have similar limiting distances.  For O3, the expected distance sensitivity of advanced LIGO to binary neutron star mergers will be $D$$\approx$120--170 Mpc, while for Virgo it will be $D$$\approx$65--85 Mpc \citep{lvk,lvknew} \footnote{These distance ranges are the volume- and orientation-averaged distance at which a CBC of a given mass yields a matched filter signal-to-noise ratio (SNR) of 8 in a single detector.}.  
%Not only does DLT40 have the necessary depth to respond to LVC triggers in O3, but we can observe the relevant galaxies within hours, facilitating the fast identification of associated transients.  As a simple demonstration, we use the 2D sky localizations of two events from O2 -- GW170817 (33 deg$^2$; 90\% confidence level) and GW170814 (97 deg$^2$; 90\% confidence level) -- as representative of three-detector events in O3, and calculate how long DLT40 would take to observe it's galaxy-targeted program if these 2D sky localizations were placed at varying distances.
If we assume that all kilonovae are as bright as AT2018gfo/DLT17ck and neglect that any potential galaxy catalog is incomplete, the current \ac{DLT40} observing strategy could detect all kilonovae in the Virgo volume during the O3 run.  In the end, rather than going out to the full $\approx$80 Mpc distance horizon of Virgo for O3, we chose a distance at the lower end of their expected sensitivity ($D$$\approx$65 Mpc), which was more practical for the purposes of gathering template images of all the necessary galaxy fields.

\subsection{DLT40 plans for O3}\label{sec:o3}

The DLT40 survey is continually upgrading its hardware and software, and we have been making special plans for the LVC O3 run -- we summarize these improvements here.

%We designed and are in the stage of applying our DLT40 plans for the upcoming LVC O3 run, as following:

\begin{enumerate}

\item As shown in Section~\ref{sec:up}, the depth and cadence of DLT40 is suitable for finding typical kilonovae out to the horizon of the Virgo detector during O3.  We are thus gathering template images of galaxies out to $D$$\approx$65 Mpc (drawn from the GWGC, with Dec$<$+20 deg, $M_B$$<$$-$18 mag and $A_V$$<$0.5 mag, as with the rest of the standard DLT40 search) in preparation for following up merger events with a neutron star in O3.

\item The DLT40 team has recently added a second identical telescope to its SN search, located at Meckering Observatory in western Australia.  This telescope will be utilized on an equal footing to the original DLT40 telescope at CTIO, and makes our GW counterpart search more robust to weather and instrument problems.  It also decreases our response time to any given event, given our distributed longitudinal coverage.  DLT40 is also in the process of incorporating a third telescope in Alberta, Canada, which will likely be operational near the currently planned start of O3 (April 2019). This telescope will provide northern hemisphere coverage for the DLT40 EM counterpart search.

\item Several recent improvements have been made to the overall DLT40 pipeline.  Chief among them has been the adoption of a pixel-based, random forest machine learning algorithm used for scoring incoming SN candidates in real time \citep[based on the algorithm used by the Pan-STARRS1 survey;][]{Wright15}.  This algorithm has drastically shrunk the number of transient candidates that must be inspected in a given night's worth of data ($\sim$10s instead of $\sim$1000s).  In addition, very strong transient candidates with high machine learning scores trigger automated email alerts to the DLT40 team, which can respond with follow-up imaging on either DLT40 telescope within minutes.

\item The DLT40 GW follow-up code will respond autonomously to incoming LVC alerts, inserting the appropriate galaxy targets into the DLT40 scheduler at high priority in real time.  It will also alert the team if a counterpart candidate appears in a selected field (see point (iii) above).  DLT40 will respond to all burst alerts in O3 (which will not have distance information) as well as all CBC alerts which have a cumulative probability of $>$10\% of being within $D$$<$65 Mpc (we intend to follow all nearby CBC alerts. In case the trigger rate is relatively high in O3, e.g. several sources within one week, we would manually evaluate them with the false alarm rate and the probability of which contains a neutron star).

\end{enumerate}

\section{Summary}
\label{sec:discussion}
We have presented the GW follow-up strategy and results for the DLT40 survey during the O2 observing run of Advanced LIGO and Virgo.  DLT40 employed a galaxy-targeted search which combined the GW localization region and the galaxy catalog employed by the DLT40 SN search to prioritize targets for follow-up.  This strategy bore fruit for GW170817, the first detected binary neutron star merger, as DLT40 co-discovered the associated kilonova AT2017gfo/SSS17a/DLT17ck \citep{dlt17ck}.  In all, we present follow-up observations of ten out of the sixteen O2 GW triggers.  Of these ten, three events were ultimately verified as {\it bona fide} GW events.

Finally, in Section~\ref{sec:future} we discussed the DLT40 team's follow-up plans for the upcoming O3 detector run.  Two additional, identical telescopes will be added in Australia and Canada, providing longitudinal and northern hemisphere coverage.  Automated response to new GW events with allow for immediate scheduling of high-priority galaxy fields on one of DLT40's search telescopes, and a new  machine learning algorithm will cleanly identify new transient events and notify the team within minutes if a strong optical counterpart candidate is found.  Further, artificial star experiments indicate that DLT40 will be sensitive to kilonovae out to the neutron star merger horizon of Virgo during O3, and we will respond to any appropriate event out to $D$$\approx$65 Mpc.  Small robotic telescope systems, coupled with smart software, will continue to have an out-sized impact on multi-messenger astrophysics in the years ahead.

\section*{Acknowledgements}
SY would like to thank Marica Branchesi and He Gao for helpful discussions and comments. 
SY acknowledges from the China Scholarship Council $\#$201506040044, and the supported by the PRIN-INAF 2016 with the project "Towards the SKA and CTA era: discovery, localization, and physics of transient source".
Research by DJS is supported by NSF grants AST-1821967, 1821987, 1813708 and 1813466.  
Research by SV is supported by NSF grant AST-1813176.
The \ac{DLT40} web-pages structure were developed at the Aspen Center for Physics, which is supported by National Science Foundation grant PHY-1066293. 
AC acknowledges support from the NSF award $\#$1455090 "CAREER: Radio and gravitational-wave emission from the largest explosions since the Big Bang".

%%%%%%%%%%%%%%%%%%%% REFERENCES %%%%%%%%%%%%%%%%%%
\bibliography{sheng_DS}

\clearpage
%%%%%%%%%%%%%%%%%%%%%%%%%%%%%%%%%%%%%%%%%%%%%%%%%%
%%%%%%%%%%%%%%%%% APPENDICES %%%%%%%%%%%%%%%%%%%%%
\appendix
%\onecolumn
\section{DLT40 GW Follow-up Observing logs for the LVC O2 run}
Table \ref{G268556}, \ref{G270580}, \ref{G274296}, \ref{G275404_1}, \ref{G275404_2}, \ref{G275697_1}, \ref{G275697_2}, \ref{G277583}, \ref{G284239}, \ref{GW170814}, \ref{GRB170817a}, \ref{GW170817}, \ref{G299232} present the DLT40 galaxies that were selected, activated and followed by us during the second observing run of the LVC. 
In each of the presented galaxy lists, we display target galaxy information \citep[from the GWGC;][]{GWGC} including the galaxy name, equatorial coordinates, luminosity distance, and absolute magnitude in B and K band.  We also include details of the observations, including the observing window, and the score estimated by our DLT40 galaxy priorization algorithm.  These information was also presented in the DLT40 team's O2 GCNs.

%show the galaxy informations (from the GWGC catalogue), including their name, equatorial coordinates, luminosity distance, absolute magnitude in B and K band, together with DLT40 informations, including the observing window, and the score estimated by our DLT40 galaxy priorization algorithm. These informations were also reported in the DLT40 GCNs correspondingly.

%%%%%%%%%%%%%%%%G268556
\startlongtable
\begin{deluxetable*}{cccccccc}
\tablecaption{Galaxies observed after trigger GW170104/G268556.  Table columns provide the galaxy name, coordinates, distance, B and K-band magnitude, observing window and the score from our ranking algorithm. The cumulative score is 0.357. \label{G268556}}
\tablehead{
\colhead{Name} & \colhead{RA(J2000)} & \colhead{DEC(J2000)} & \colhead{Dist(Mpc)} & \colhead{B(mag)}  & \colhead{K(mag)} & \colhead{OBS\_WINDOW(JD)} & \colhead{SCORE}
}
\colnumbers
\startdata
NGC7507 & 348.03 & -28.54 & 25.00 & -20.77 & -24.70 & 2457759.33-2457766.29 & 2.033$\times10^{-1}$ \\ 
\hline 
NGC7755 & 356.97 & -30.52 & 27.92 & -20.63 & -23.26 & 2457759.33-2457766.29 & 6.838$\times10^{-2}$ \\ 
\hline 
NGC1291 & 49.33 & -41.11 & 10.38 & -20.66 & -24.42 & 2457759.33-2457766.29 & 1.823$\times10^{-2}$ \\ 
\hline 
NGC1448 & 56.13 & -44.64 & 15.35 & -20.30 & -23.27 & 2457759.33-2457766.29 & 1.013$\times10^{-2}$ \\ 
\hline 
NGC1365 & 53.40 & -36.14 & 17.95 & -21.39 & -24.90 & 2457759.33-2457766.29 & 8.132$\times10^{-3}$ \\ 
\hline 
NGC2263 & 99.62 & -24.85 & 32.36 & -20.75 & -23.41 & 2457759.33-2457766.29 & 5.403$\times10^{-3}$ \\ 
\hline 
NGC1411 & 54.69 & -44.10 & 19.77 & -19.78 & -23.33 & 2457759.33-2457766.29 & 5.345$\times10^{-3}$ \\ 
\hline 
ESO557-009 & 99.68 & -20.29 & 32.08 & -20.52 & -21.91 & 2457759.33-2457766.29 & 4.550$\times10^{-3}$ \\
\enddata
\end{deluxetable*}

%%%%%%%%%%%%%%%%G270580
\startlongtable
\begin{deluxetable*}{cccccccc}
\tablecaption{Galaxies observed after trigger G270580.  Table columns provide the galaxy name, coordinates, distance, B and K-band magnitude, observing window and the score from our ranking algorithm. The cumulative score is 0.106. \label{G270580}}
\tablehead{
\colhead{Name} & \colhead{RA(J2000)} & \colhead{DEC(J2000)} & \colhead{Dist(Mpc)} & \colhead{B(mag)}  & \colhead{K(mag)} & \colhead{OBS\_WINDOW(JD)} & \colhead{SCORE}
}
\colnumbers
\startdata
NGC4419 & 186.74 & 15.05 & 13.49 & -19.24 & -22.91 & 2457991.48-2458005.21 & 2.703$\times10^{-2}$ \\ 
\hline 
NGC4216 & 183.98 & 13.15 & 14.13 & -20.58 & -24.23 & 2457991.48-2458005.21 & 1.633$\times10^{-2}$ \\ 
\hline 
IC5201 & 335.24 & -46.03 & 9.20 & -18.78 & -19.65 & 2457991.48-2458005.21 & 9.84$\times10^{-3}$ \\ 
\hline 
NGC4571 & 189.23 & 14.22 & 14.93 & -19.12 & -22.35 & 2457991.48-2458005.21 & 9.603$\times10^{-3}$ \\ 
\hline 
NGC4498 & 187.91 & 16.85 & 14.06 & -18.54 & -21.08 & 2457991.48-2458005.21 & 9.24$\times10^{-3}$ \\ 
\hline 
NGC4192 & 183.45 & 14.90 & 13.61 & -20.62 & -23.78 & 2457991.48-2458005.21 & 5.19$\times10^{-3}$ \\ 
\hline 
NGC4178 & 183.19 & 10.87 & 14.00 & -19.37 & -21.16 & 2457991.48-2458005.21 & 5.118$\times10^{-3}$ \\ 
\hline 
NGC1546 & 63.65 & -56.06 & 13.87 & -18.21 & -22.66 & 2457991.48-2458005.21 & 4.41$\times10^{-3}$ \\
\enddata
\end{deluxetable*}

%%%%%%%%%%%%%%%%G274296
\startlongtable
\begin{deluxetable*}{cccccccc}
\tablecaption{Galaxies observed after trigger G274296.  Table columns provide the galaxy name, coordinates, distance, B and K-band magnitude, observing window and the score from our ranking algorithm. The cumulative score is 0.704. \label{G274296}}
\tablehead{
\colhead{Name} & \colhead{RA(J2000)} & \colhead{DEC(J2000)} & \colhead{Dist(Mpc)} & \colhead{B(mag)}  & \colhead{K(mag)} & \colhead{OBS\_WINDOW(JD)} & \colhead{SCORE}
}
\colnumbers
\startdata
IC4837A & 288.82 & -54.13 & 33.42 & -21.15 & -24.47 & 2457802.38-2457825.37 & 4.803$\times10^{-2}$ \\ 
\hline 
IC4837 & 288.81 & -54.67 & 33.42 & -20.72 & -20.38 & 2457802.38-2457825.37 & 3.770$\times10^{-2}$ \\ 
\hline 
IC4839 & 288.89 & -54.63 & 35.11 & -20.02 & -23.48 & 2457802.38-2457825.37 & 3.770$\times10^{-2}$ \\ 
\hline 
NGC3041 & 148.28 & 16.68 & 23.77 & -19.86 & -23.14 & 2457802.38-2457825.37 & 3.738$\times10^{-2}$ \\ 
\hline 
NGC6788 & 291.71 & -54.95 & 34.56 & -21.09 & -24.17 & 2457802.38-2457825.37 & 3.607$\times10^{-2}$ \\ 
\hline 
NGC2698 & 133.90 & -3.18 & 24.94 & -18.89 & -23.14 & 2457802.38-2457825.37 & 3.437$\times10^{-2}$ \\ 
\hline 
NGC2708 & 134.03 & -3.36 & 26.55 & -19.15 & -23.24 & 2457802.38-2457825.37 & 3.437$\times10^{-2}$ \\ 
\hline 
NGC2697 & 133.75 & -2.99 & 24.17 & -18.59 & -22.02 & 2457802.38-2457825.37 & 3.041$\times10^{-2}$ \\
\enddata
\end{deluxetable*}

%%%%%%%%%%%%%%%%G275404
\startlongtable
\begin{deluxetable*}{cccccccc}
\tablecaption{Galaxies observed after trigger G275404 (with the preliminary bayestar localization map).  Table columns provide the galaxy name, coordinates, distance, B and K-band magnitude, observing window and the score from our ranking algorithm. The cumulative sore is 0.621. \label{G275404_1}}
\tablehead{
\colhead{Name} & \colhead{RA(J2000)} & \colhead{DEC(J2000)} & \colhead{Dist(Mpc)} & \colhead{B(mag)}  & \colhead{K(mag)} & \colhead{OBS\_WINDOW(JD)} & \colhead{SCORE}
}
\colnumbers
\startdata
NGC3923 & 177.76 & -28.81 & 22.91 & -20.97 & -25.30 & 2457810.63-2457821.57 & 4.229$\times10^{-2}$ \\ 
\hline 
NGC3904 & 177.30 & -29.28 & 28.31 & -20.53 & -24.58 & 2457810.63-2457821.57 & 3.748$\times10^{-2}$ \\ 
\hline 
ESO440-027 & 178.35 & -28.55 & 18.53 & -18.93 & -21.74 & 2457810.63-2457821.57 & 3.080$\times10^{-2}$ \\ 
\hline 
ESO440-011 & 177.19 & -28.29 & 21.88 & -19.21 & -18.05 & 2457810.63-2457821.57 & 2.991$\times10^{-2}$ \\ 
\hline 
IC3010 & 181.99 & -30.34 & 25.87 & -19.21 & -22.86 & 2457810.63-2457821.57 & 2.848$\times10^{-2}$ \\ 
\hline 
IC3005 & 181.81 & -30.02 & 24.32 & -19.01 & -21.96 & 2457810.63-2457821.57 & 2.618$\times10^{-2}$ \\ 
\hline 
NGC4105 & 181.67 & -29.76 & 26.55 & -20.32 & -24.60 & 2457810.63-2457821.57 & 2.421$\times10^{-2}$ \\ 
\hline 
ESO441-014 & 182.56 & -30.08 & 24.67 & -18.60 & -19.56 & 2457810.63-2457821.57 & 2.245$\times10^{-2}$ \\
\enddata
\end{deluxetable*}

\startlongtable
\begin{deluxetable*}{cccccccc}
\tablecaption{Galaxies observed after trigger G275404 (with the updated LALInference localization map).  Table columns provide the galaxy name, coordinates, distance, B and K-band magnitude, observing window and the score from our ranking algorithm. The cumulative sore is 0.243. \label{G275404_2}}
\tablehead{
\colhead{Name} & \colhead{RA(J2000)} & \colhead{DEC(J2000)} & \colhead{Dist(Mpc)} & \colhead{B(mag)}  & \colhead{K(mag)} & \colhead{OBS\_WINDOW(JD)} & \colhead{SCORE}
}
\colnumbers
\startdata
NGC3923 & 177.76 & -28.81 & 22.91 & -20.97 & -25.30 & 2457821.57-2457825.39 & 1.209$\times10^{-2}$ \\ 
\hline 
NGC3904 & 177.30 & -29.28 & 28.31 & -20.53 & -24.58 & 2457821.57-2457825.39 & 1.112$\times10^{-2}$ \\ 
\hline 
ESO380-006 & 183.89 & -35.63 & 38.93 & -21.31 & -24.79 & 2457821.57-2457825.39 & 8.775$\times10^{-3}$ \\ 
\hline 
NGC4603 & 190.23 & -40.98 & 33.11 & -21.20 & -24.23 & 2457821.57-2457825.39 & 8.708$\times10^{-3}$ \\ 
\hline 
NGC4696 & 192.21 & -41.31 & 35.48 & -21.47 & -25.61 & 2457821.57-2457825.39 & 7.815$\times10^{-3}$ \\ 
\hline 
IC3253 & 185.94 & -34.62 & 34.67 & -21.00 & -23.59 & 2457821.57-2457825.39 & 7.541$\times10^{-3}$ \\ 
\hline 
NGC4105 & 181.67 & -29.76 & 26.55 & -20.32 & -24.60 & 2457821.57-2457825.39 & 7.295$\times10^{-3}$ \\ 
\hline 
IC0764 & 182.56 & -29.74 & 21.18 & -20.13 & -21.92 & 2457821.57-2457825.39 & 6.613$\times10^{-3}$ \\
\enddata
\end{deluxetable*}

%%%%%%%%%%%%%%%%G275697
\startlongtable
\begin{deluxetable*}{cccccccc}
\tablecaption{Galaxies observed after trigger G275697 (with the preliminary bayestar localization map). Table columns provide the galaxy name, coordinates, distance, B and K-band magnitude, observing window and the score from our ranking algorithm. The cumulative score is 0.681. \label{G275697_1}}
\tablehead{
\colhead{Name} & \colhead{RA(J2000)} & \colhead{DEC(J2000)} & \colhead{Dist(Mpc)} & \colhead{B(mag)}  & \colhead{K(mag)} & \colhead{OBS\_WINDOW(JD)} & \colhead{SCORE}
}
\colnumbers
\startdata
NGC3742 & 173.89 & -37.96 & 38.61 & -20.57 & -24.29 & 2457812.34-2457820.41  & 4.087$\times10^{-2}$ \\ 
\hline 
NGC3749 & 173.97 & -38.00 & 38.02 & -20.80 & -24.20 & 2457812.34-2457820.41  & 3.951$\times10^{-2}$ \\ 
\hline 
ESO320-026 & 177.46 & -38.78 & 38.02 & -20.92 & -24.22 & 2457812.34-2457820.41  & 3.305$\times10^{-2}$ \\ 
\hline 
ESO320-004 & 173.68 & -38.25 & 38.02 & -19.46 & -21.31 & 2457812.34-2457820.41  & 3.203$\times10^{-2}$ \\ 
\hline 
ESO320-027 & 177.60 & -38.65 & 37.64 & -19.93 & -20.41 & 2457812.34-2457820.41  & 3.092$\times10^{-2}$ \\ 
\hline 
ESO320-024 & 177.36 & -38.83 & 38.89 & -19.55 & -21.80 & 2457812.34-2457820.41  & 3.087$\times10^{-2}$ \\ 
\hline 
NGC3783 & 174.76 & -37.74 & 38.02 & -20.50 & -24.25 & 2457812.34-2457820.41  & 2.898$\times10^{-2}$ \\ 
\hline 
ESO320-030 & 178.30 & -39.13 & 38.02 & -20.25 & -23.69 & 2457812.34-2457820.41  & 2.888$\times10^{-2}$ \\
\enddata
\end{deluxetable*}

\startlongtable
\begin{deluxetable*}{cccccccc}
\tablecaption{Galaxies observed after trigger G275697 (with the updated LALInference localization map).  Table columns provide the galaxy name, coordinates, distance, B and K-band magnitude, observing window and the score from our ranking algorithm. The cumulative sore is 0.574. \label{G275697_2}}
\tablehead{
\colhead{Name} & \colhead{RA(J2000)} & \colhead{DEC(J2000)} & \colhead{Dist(Mpc)} & \colhead{B(mag)}  & \colhead{K(mag)} & \colhead{OBS\_WINDOW(JD)} & \colhead{SCORE}
}
\colnumbers
\startdata
NGC3742 & 173.89 & -37.96 & 38.61 & -20.57 & -24.29 & 2457820.41-2457825.39 & 1.823$\times10^{-2}$ \\ 
\hline 
NGC3749 & 173.97 & -38.00 & 38.02 & -20.80 & -24.20 & 2457820.41-2457825.39 & 1.761$\times10^{-2}$ \\ 
\hline 
NGC3557B & 167.38 & -37.35 & 36.98 & -20.06 & -23.75 & 2457820.41-2457825.39 & 1.554$\times10^{-2}$ \\ 
\hline 
NGC3573 & 167.83 & -36.88 & 37.18 & -19.87 & -24.02 & 2457820.41-2457825.39 & 1.538$\times10^{-2}$ \\ 
\hline 
ESO320-004 & 173.68 & -38.25 & 38.02 & -19.46 & -21.31 & 2457820.41-2457825.39 & 1.495$\times10^{-2}$ \\ 
\hline 
ESO320-026 & 177.46 & -38.78 & 38.02 & -20.92 & -24.22 & 2457820.41-2457825.39 & 1.454$\times10^{-2}$ \\ 
\hline 
NGC3564 & 167.65 & -37.55 & 36.98 & -20.08 & -23.89 & 2457820.41-2457825.39 & 1.409$\times10^{-2}$ \\ 
\hline 
ESO320-027 & 177.60 & -38.65 & 37.64 & -19.93 & -20.41 & 2457820.41-2457825.39 & 1.399$\times10^{-2}$ \\
\enddata
\end{deluxetable*}

%%%%%%%%%%%%%%%%G277583
\startlongtable
\begin{deluxetable*}{cccccccc}
\tablecaption{Galaxies observed after trigger G277583.  Table columns provide the galaxy name, coordinates, distance, B and K-band magnitude, observing window and the score from our ranking algorithm. The cumulative score is 0.370. \label{G277583}}
\tablehead{
\colhead{Name} & \colhead{RA(J2000)} & \colhead{DEC(J2000)} & \colhead{Dist(Mpc)} & \colhead{B(mag)}  & \colhead{K(mag)} & \colhead{OBS\_WINDOW(JD)} & \colhead{SCORE}
}
\colnumbers
\startdata
NGC1407 & 55.05 & -18.58 & 28.84 & -21.65 & -25.60 & 2457826.53-2457847.24 & 1.595$\times10^{-2}$ \\ 
\hline 
NGC1400 & 54.88 & -18.69 & 26.42 & -20.17 & -24.30 & 2457826.53-2457847.24 & 1.502$\times10^{-2}$ \\ 
\hline 
NGC1393 & 54.66 & -18.43 & 23.89 & -19.49 & -22.71 & 2457826.53-2457847.24 & 1.293$\times10^{-2}$ \\ 
\hline 
NGC1421 & 55.62 & -13.49 & 25.35 & -21.11 & -23.62 & 2457826.53-2457847.24 & 1.102$\times10^{-2}$ \\ 
\hline 
NGC1625 & 69.28 & -3.30 & 36.98 & -20.44 & -20.41 & 2457826.53-2457847.24 & 8.936$\times10^{-3}$ \\ 
\hline 
PGC013646 & 55.73 & -12.92 & 25.35 & -19.62 & -22.63 & 2457826.53-2457847.24 & 8.625$\times10^{-3}$ \\ 
\hline 
IC2098 & 72.68 & -5.42 & 32.36 & -19.77 & -22.35 & 2457826.53-2457847.24 & 8.333$\times10^{-3}$ \\ 
\hline 
NGC1665 & 72.07 & -5.43 & 33.10 & -19.32 & -23.34 & 2457826.53-2457847.24 & 6.994$\times10^{-3}$ \\
\enddata
\end{deluxetable*}

%%%%%%%%%%%%%%%%G284239
\startlongtable
\begin{deluxetable*}{cccccccc}
\tablecaption{Galaxies observed after trigger G284239.  Table columns provide the galaxy name, coordinates, distance, B and K-band magnitude, observing window and the score from our ranking algorithm. The cumulative score is 0.846. \label{G284239}}
\tablehead{
\colhead{Name} & \colhead{RA(J2000)} & \colhead{DEC(J2000)} & \colhead{Dist(Mpc)} & \colhead{B(mag)}  & \colhead{K(mag)} & \colhead{OBS\_WINDOW(JD)} & \colhead{SCORE}
}
\colnumbers
\startdata
NGC6861 & 301.83 & -48.37 & 28.05 & -20.29 & -24.53 & 2457877.13-2457892.23 & 6.452$\times10^{-2}$ \\ 
\hline 
ESO233-035 & 302.36 & -48.28 & 33.97 & -18.57 & -22.23 & 2457877.13-2457892.23 & 6.320$\times10^{-2}$ \\ 
\hline 
NGC6861D & 302.08 & -48.21 & 33.42 & -18.90 & -23.42 & 2457877.13-2457892.23 & 6.280$\times10^{-2}$ \\ 
\hline 
NGC6868 & 302.48 & -48.38 & 26.79 & -20.45 & -24.82 & 2457877.13-2457892.23 & 6.110$\times10^{-2}$ \\ 
\hline 
NGC6870 & 302.55 & -48.29 & 33.42 & -19.90 & -23.67 & 2457877.13-2457892.23 & 6.030$\times10^{-2}$ \\ 
\hline 
NGC6861E & 302.76 & -48.69 & 33.90 & -18.55 & -21.04 & 2457877.13-2457892.23 & 4.919$\times10^{-2}$ \\ 
\hline 
NGC6851 & 300.89 & -48.28 & 36.14 & -20.13 & -24.02 & 2457877.13-2457892.23 & 4.075$\times10^{-2}$ \\ 
\hline 
NGC6384 & 263.10 & 7.06 & 25.94 & -21.26 & -24.54 & 2457877.13-2457892.23 & 2.247$\times10^{-2}$ \\
\enddata
\end{deluxetable*}

%%%%%%%%%%%%%%%%GW170814/G297595
\startlongtable
\begin{deluxetable*}{cccccccc}
\tablecaption{Galaxies observed after trigger GW170814/G297595.  Table columns provide the galaxy name, coordinates, distance, B and K-band magnitude, observing window and the score from our ranking algorithm. The cumulative score is 0.550. \label{GW170814}}
\tablehead{
\colhead{Name} & \colhead{RA(J2000)} & \colhead{DEC(J2000)} & \colhead{Dist(Mpc)} & \colhead{B(mag)}  & \colhead{K(mag)} & \colhead{OBS\_WINDOW(JD)} & \colhead{SCORE}
}
\colnumbers
\startdata
ESO300-014 & 47.41 & -41.03 & 10.86 & -18.32 & -16.32 & 2457980.28-2457982.25 & 4.898$\times10^{-2}$ \\ 
\hline 
NGC1255 & 48.38 & -25.73 & 17.70 & -19.83 & -22.86 & 2457980.28-2457982.25 & 4.695$\times10^{-2}$ \\ 
\hline 
ESO481-018 & 49.64 & -25.84 & 18.53 & -18.28 & -20.64 & 2457980.28-2457982.25 & 4.544$\times10^{-2}$ \\ 
\hline 
NGC0986 & 38.39 & -39.05 & 23.12 & -20.16 & -24.04 & 2457980.28-2457982.25 & 4.508$\times10^{-2}$ \\ 
\hline 
NGC1201 & 46.03 & -26.07 & 19.33 & -20.01 & -23.76 & 2457980.28-2457982.25 & 4.045$\times10^{-2}$ \\ 
\hline 
NGC1302 & 49.96 & -26.06 & 19.59 & -20.06 & -23.63 & 2457980.28-2457982.25 & 4.039$\times10^{-2}$ \\ 
\hline 
NGC0908 & 35.77 & -21.23 & 15.78 & -20.53 & -23.76 & 2457980.28-2457982.25 & 2.868$\times10^{-2}$ \\ 
\hline 
PGC013359 & 54.31 & 1.86 & 22.59 & -19.15 & -20.80 & 2457980.28-2457982.25 & 2.490$\times10^{-2}$ \\
\enddata
\end{deluxetable*}

%%%%%%%%%%%%%%%% Fermi
\startlongtable
\begin{deluxetable*}{ccccccc}
\tablecaption{Galaxies observed after trigger GRB170817a.  Table columns provide the galaxy name, coordinates, distance, B and K-band magnitude and the observing window. For the GRB trigger, we selected the 31 most luminous galaxies in the 90\% Fermi region \citep{dlt17ck}. \label{GRB170817a}}
\tablehead{
\colhead{Name} & \colhead{RA(J2000)} & \colhead{DEC(J2000)} & \colhead{Dist(Mpc)} & \colhead{B(mag)}  & \colhead{K(mag)} & \colhead{OBS\_WINDOW(JD)} 
}
\colnumbers
\startdata
NGC3749 & 173.97 & -38.00 & 38.02 & -20.80 & -24.20 & 2457980.28-2457982.25  \\ 
\hline 
NGC3269 & 157.49 & -35.22 & 36.40 & -20.86 & -23.75 & 2457980.28-2457982.25  \\ 
\hline 
NGC3273 & 157.62 & -35.61 & 36.98 & -18.99 & -23.91 & 2457980.28-2457982.25  \\ 
\hline 
NGC3281C & 158.25 & -34.89 & 37.29 & -18.79 & -22.62 & 2457980.28-2457982.25 \\ 
\hline 
ESO268-003 & 186.32 & -43.45 & 37.92 & -18.95 & -23.22 & 2457980.28-2457982.25 \\ 
\hline 
ESO443-042 & 195.87 & -29.83 & 33.42 & -19.69 & -22.94 & 2457980.28-2457982.25  \\ 
\hline 
ESO440-004 & 176.42 & -28.37 & 22.93 & -18.82 & -18.47 & 2457980.28-2457982.25  \\ 
\hline 
IC4180 & 196.74 & -23.92 & 33.65 & -19.98 & -23.04 & 2457980.28-2457982.25 \\
\enddata
\end{deluxetable*}

%%%%%%%%%%%%%%%%GW170817/G298048
\startlongtable
\begin{deluxetable*}{ccccccccc}
\tablecaption{Galaxies observed after trigger GW170817/G298048.  Table columns provide the galaxy name, coordinates, distance, B and K-band magnitude, observing window and the score from our ranking algorithm. It's shown that NGC4993, which is the host galaxy of the kilonova AT17fgo, is ranked by our algorithm as the second in the list, see column 'score'. After taking account of the distance distribution from GW estimation, NGC4993 is the highest ranking galaxy, shown as column 'score2'. The galaxy list is ranked with 'score2'. \label{GW170817}}
\tablehead{
\colhead{Name} & \colhead{RA(J2000)} & \colhead{DEC(J2000)} & \colhead{Dist(Mpc)} & \colhead{B(mag)}  & \colhead{K(mag)} & \colhead{OBS\_WINDOW(JD)} & \colhead{SCORE} & \colhead{SCORE2}
}
\colnumbers
\startdata
{\color{red} NGC4993} & {\color{red} 197.45} & {\color{red} -23.38} & {\color{red} 39.99} & {\color{red} -20.20} & {\color{red} -23.42} & {\color{red} 2457983.41-2457985.17} & {\color{red} 1.195$\times10^{-1}$} & {\color{red} 1.615$\times10^{-1}$} \\ 
\hline 
IC4197 & 197.02 & -23.80 & 34.10 & -20.24 & -23.39 & 2457983.41-2457985.17  & 1.346$\times10^{-1}$ & 1.386$\times10^{-1}$ \\ 
\hline 
ESO508-019 & 197.47 & -24.24 & 38.55 & -19.33 & -21.16 & 2457983.41-2457985.17  &  1.060$\times10^{-1}$ & 1.311$\times10^{-1}$ \\ 
\hline 
NGC4968 & 196.77 & -23.68 & 33.42 & -19.44 & -23.14 & 2457983.41-2457985.17  &  1.145$\times10^{-1}$ & 1.103$\times10^{-1}$ \\ 
\hline 
ESO508-024 & 197.69 & -23.87 & 33.42 & -19.98 & -21.59 & 2457983.41-2457985.17  &  9.862$\times10^{-2}$ & 1.021$\times10^{-1}$ \\ 
\hline 
IC4180 & 196.74 & -23.92 & 33.65 & -19.98 & -23.04& 2457983.41-2457985.17  &  9.361$\times10^{-2}$ & 9.230$\times10^{-2}$ \\ 
\hline 
ESO508-015 & 197.33 & -24.38 & 33.42 & -18.77 & -18.79 & 2457983.41-2457985.17  &  8.194$\times10^{-2}$ & 7.894$\times10^{-2}$ \\ 
\hline 
ESO508-004 & 196.72 & -22.84 & 33.96 & -18.33 & -18.75 & 2457983.41-2457985.17  &  7.466$\times10^{-2}$ & 7.586$\times10^{-2}$ \\
\enddata
\end{deluxetable*}

%%%%%%%%%%%%%%%%G299232
\startlongtable
\begin{deluxetable*}{cccccccc}
\tablecaption{Galaxies observed after trigger G299232.  Table columns provide the galaxy name, coordinates, distance, B and K-band magnitude, observing window and the score from our ranking algorithm. The cumulative score is 0.775. \label{G299232}}
\tablehead{
\colhead{Name} & \colhead{RA(J2000)} & \colhead{DEC(J2000)} & \colhead{Dist(Mpc)} & \colhead{B(mag)}  & \colhead{K(mag)} & \colhead{OBS\_WINDOW(JD)} & \colhead{SCORE} 
}
\colnumbers
\startdata
NGC0772 & 29.83 & 19.01 & 32.36 & -21.78 & -25.35 & 2457991.48-2458005.21 & 5.598$\times10^{-2}$ \\ 
\hline 
NGC0524 & 21.20 & 9.54 & 23.99 & -20.53 & -24.74 & 2457991.48-2458005.21 & 5.373$\times10^{-2}$ \\ 
\hline 
NGC0770 & 29.81 & 18.95 & 32.36 & -18.75 & -22.27 & 2457991.48-2458005.21 & 5.272$\times10^{-2}$ \\ 
\hline 
NGC0518 & 21.07 & 9.33 & 30.31 & -19.35 & -22.61 & 2457991.48-2458005.21 & 5.084$\times10^{-2}$ \\ 
\hline 
NGC0525 & 21.22 & 9.70 & 22.32 & -18.32 & -21.14 & 2457991.48-2458005.21 & 5.020$\times10^{-2}$ \\ 
\hline 
NGC0516 & 21.03 & 9.55 & 29.87 & -18.82 & -21.86 & 2457991.48-2458005.21 & 4.811$\times10^{-2}$ \\ 
\hline 
NGC0532 & 21.32 & 9.26 & 29.92 & -19.73 & -23.55 & 2457991.48-2458005.21 & 4.614$\times10^{-2}$ \\ 
\hline 
IC0101 & 21.04 & 9.93 & 30.71 & -18.40 & -20.56 & 2457991.48-2458005.21 & 4.544$\times10^{-2}$ \\
\enddata
\end{deluxetable*}
\end{document}